\documentclass[%
reprint,
 amsmath,amssymb,
 aps,
prb,
]{revtex4-2}

\usepackage{graphicx}% Include figure files
\usepackage{dcolumn}% Align table columns on decimal point
\usepackage{bm}% bold math
\usepackage[colorlinks=true,citecolor=blue,linkcolor=red,urlcolor=blue]{hyperref}
\begin{document}

\title{Smoking-gun signatures of non-Markovianity of a superconducting qubit}% Work in progress title

\author{Bal\'azs Gul\'acsi}
\email{balazs.gulacsi@uni-konstanz.de}
\author{Guido Burkard}
 \email{guido.burkard@uni-konstanz.de}
\affiliation{%
 Department of Physics, University of Konstanz, 78457 Konstanz, Germany
}%

\date{\today}% It is always \today, today,

\begin{abstract}
We describe temporally correlated noise processes that influence the idle evolution of a superconducting transmon qubit.
To model the composite qubit-environment system we use quantum circuit theory, and we show how a circuit Hamiltonian can be derived for transverse noise affecting the qubit. 
Based on the time-convolutionless projection operator method, we construct a time-local master equation which, when transformed to its canonical Lindblad form, exhbitis a decay rate that is negative at all times, corresponding to eternally non-Markovian dynamics. 
By expressing the solution of the master equation in the Kraus representation, we identify two crucial non-Markovian phenomena: periodic revivals of coherence, and the appearance of additional frequencies far from the qubit frequency in the precession of the qubit state.
When a single qubit gate acts on the qubit state, these extra frequency terms rotate undesirably and they effectively act as the memory of the state prior to the rotation around the Bloch sphere.
\end{abstract}

\maketitle

%\tableofcontents

\section{\label{sec:level1}Introduction}

Electrical circuits containing superconducting elements are currently one of the leading platforms on the road for reaching fault-tolerant quantum computation. In these circuits, Josepshon junctions are combined with capacitors and inductors, effectively acting as non-linear circuit elements which transform the circuit into an anharmonic oscillator behaving as an artificial atom \cite{Gambetta2017,Wendin2017,Martinis2020}.
Based on the varying relative strengths of the characteristic circuit energies associated with the inductance $L$, capacitance $C$, and Josephson elements, these circuit constructions allow for highly diverse types of superconducting qubits \cite{scqubit1}, such as the transmon \cite{Transmon1,Transmon2,xmon}, quantronium \cite{quantronium}, fluxonium \cite{fluxonium}, and others \cite{more1,Transmon3,more2,more3}.
Of late, as quantum computing technologies have entered the noisy intermediate-scale quantum era \cite{nisq}, the transmon qubit has emerged as the dominant candidate among the superconducting qubits \cite{qsup1}.

The transmon is essentially a capacitively shunted charge qubit. A charge qubit or Cooper-pair box \cite{Bouchiat1998,Nakamura1999} consists of a small superconducting island connected to a large superconducting reservoir via a Josephson junction. Information then is encoded within the states which represent the presence or absence of excess Cooper pairs on the small island.
The tunneling of Cooper pairs in these architectures is governed by two energy scales: the charging energy $E_C\sim e^2/C$ and the Josephson energy $E_J$. The transmon is operated in the parameter regime where $E_J/E_C\gg1$, which is highly beneficial owing to the exponentially suppressed sensitivity to charge noise \cite{Transmon1}. Nevertheless, transmons are still vulnerable to dephasing and relaxation caused by interactions with various noise sources. 
For example, noise sources consist of two-level-systems residing at the material interfaces and surfaces \cite{Dielectric,Dial2016}, cosmic rays and background radiation producing non-equilibrium quasiparticles by breaking Cooper pairs \cite{Vepslinen2020,Wilen2021,McEwen2021}. Furthermore, the strong interaction of transmons with the wiring of the electrical circuit, in which they are embedded, facilitates their integration with fast control and readout.
On the other hand, these couplings also imply a significant interaction between the qubits and their electromagnetic environment \cite{Transmon4}, leading to another source of noise.

As environmental effects inevitably delete existing quantum coherence necessary for quantum computation, it is essential to perform quantum error correction \cite{Unruh1995,Shor1995,Steane1996}. However, quantum error correction heavily relies on the properties of the noise \cite{Devitt2013}. For instance, the vast majority of quantum error-correcting codes are designed for independent error models and the presence of correlations between noise at different times and locations can ultimately cause these schemes to fail \cite{QEC1,QEC2}. 
Fortunately, due to the quantum threshold theorem, provided that noise rates remain below a certain point, error correction is theoretically possible, even in the presence of spatially and temporally correlated errors \cite{QECcorr1,Terhal2005,Aliferis2006,Preskill2013}. However, we need to be aware of which error-correcting code to use under a given set of circumstances. As a result, a deeper and more accurate characterization of noise is of paramount importance.

Here, we set out to model the influence of an electromagnetic environment on the idle time evolution of a transmon qubit. In particular, we present a calculation of the qubit dynamics in the presence of temporally correlated noise by setting up a time-local, non-Markovian master equation. A master equation is a differential equation describing the dynamical evolution of the reduced density matrix of an open quantum system. The density matrix is Hermitian and has unital trace, and the structure of the master equation must be such that these properties are preserved during the dynamics \cite{BRE02}. Previously, it was established that a trace and Hermiticity preserving time-local master equation can always be written in the well-known canonical Lindblad form \cite{Hall2014,deVega2017},
%\begin{eqnarray}
\begin{multline}
\dot\rho(t)=-i[H(t),\rho(t)]\\
+\sum_{k}\gamma_k(t)\left(L_k(t)\rho(t)L_k^\dagger(t)-\frac{1}{2}\{L_k^\dagger(t)L_k(t),\rho(t)\}\right).
\label{eq:lind}
\end{multline}
%\end{eqnarray}
Here, $\rho(t)$ is the density matrix, $H=H^\dagger$ denotes the system Hamiltonian, and the rates $\gamma_k(t)$ and jump operators $L_k(t)$ can depend explicitly on time. Since the density matrix $\rho$ is positive semidefinite, the master equation has to preserve positivity. Moreover, the dynamics have to be completely positive (CP), which is ensured whenever $\gamma_k(t)\geq0$ for all $k$ and $t$ \cite{Chruscinski2012}. As such, for time-independent non-negative coefficients and time-independent jump operators the dynamics is CP, and the generator on the right hand side of the Lindblad equation forms a Markovian semigroup \cite{Lindblad1976,GKS}.  The dynamical process described by Eq.~\eqref{eq:lind} remains Markovian as long as  $\gamma_k(t)\geq0$, which constitutes an extension of the Markovian property of dynamical processes in the time-dependent context. Accordingly, the semigroup property is replaced with CP-divisibility \cite{Rivas2014}. However, if one of the rates becomes negative at some time intervals during the time evolution, then at these time intervals CP-divisibility of the dynamical map does not hold anymore and hence the dynamics is non-Markovian.
%3. step: Explain non-Markovianity in the time local context.

The remainder of this paper is structured as follows. In Sec.~\ref{sec:model}, we describe the interaction of a transmon qubit with its electromagnetic environment using a lumped circuit model and obtain the Hamiltonian of the composite qubit-environment system via circuit theory. In Sec.~\ref{sec:TCL}, we demonstrate how to use the time-convolutionless projection technique, in order to construct the non-Markovian master equation of the open transmon system. Within Sec.~\ref{sec:TCLA} and Sec.~\ref{sec:TCLB}, we analyze the properties of the master equation and give its solution in the form of a dynamical map. In Sec.~\ref{sec:examples}, we apply our theory on two specific bath types and consider the implications of non-Markovian noise on transmon qubits. Finally, in Sec.~\ref{sec:conclusion} we summarize and conclude our work.
%4. step: final paragraph should be the structure of the paper, what we do in each section, etc.

%Here, we set out to model the influence of an electromagnetic environment on the idle time evolution of a transmon qubit. In particular, we present a calculation of the qubit dynamics in the presence of temporally correlated noise by setting up a local in time, non-Markovian master equation. 3. step: Explain non-Markovianity in the time local context.

%4. step: final paragraph should be the structure of the paper, what we do in each section, etc.

%\newpage

\begin{figure}[ht]
\includegraphics[height=4.8cm]{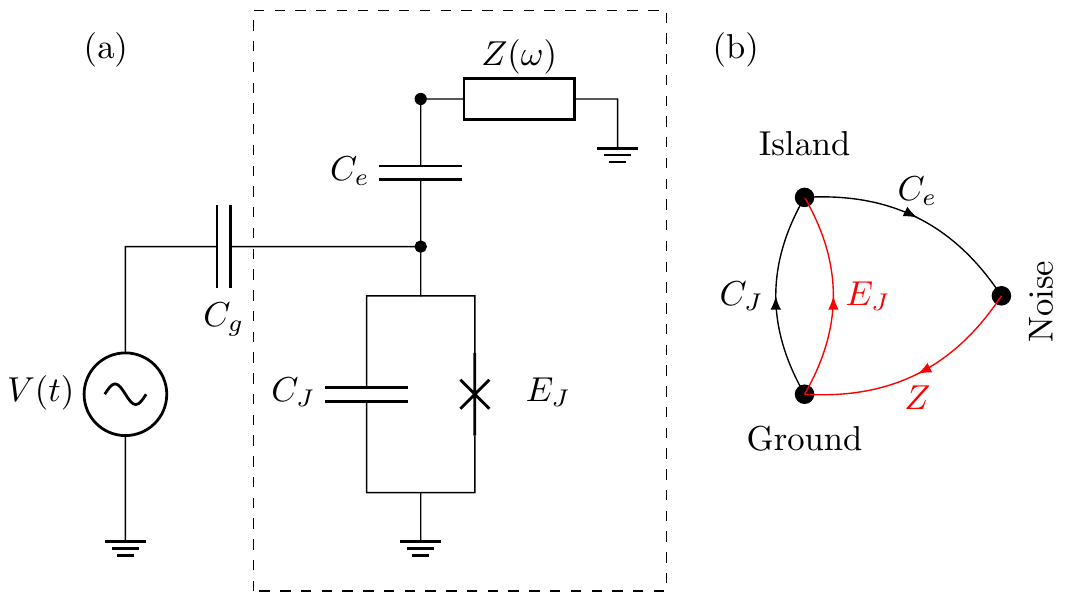}
\caption{\label{fig:circuit} Lumped circuit model of a driven transmon connected to an environment envisioned as a Caldeira-Leggett impedance. (a) The transmon consists of a capacitively shunted Josephson junction, with the sum of the intrinsic and shunt capacitances being $C_J$ and with the Josepshon energy $E_J$. The environmental impedance $Z(\omega)$ is capacitively coupled to the transmon through $C_e$. (b) Graph representation of the circuit inside the dashed rectangle in (a), with red marking the tree of the graph. }
\end{figure}

%%%%%THE CIRCUIT MODEL DESCRIPTION OF THE TRANSMON SYSTEM
%%%%%IT SHOULD END WITH THE REALIZATION THAT WE HAVE EVERYTHING FOR THE OPEN SYSTEM DESCRIPTION

\section{\label{sec:model}Model}

We model the transmon qubit and its electromagnetic environment using the lumped circuit model depicted in Fig.~\ref{fig:circuit}. We apply circuit theory \cite{Burkard2004,Burkard2005,Solgun2014,Solgun2015,Parra2019,Riwar2022} in order to obtain the Hamiltonian describing the composite transmon-environment system. One of the main advantages of this method is that the extension of the lumped model with further elements, e.g. readout resonators or more qubits, can be easily incorporated. The Hamiltonian of the circuit, in units with $\hbar=1$, reads
\begin{eqnarray}
H=4E_Cn^2-E_J\cos\varphi-n\Omega(t)+H_{Z}+\tilde\eta enB,
\label{eq:lump}
\end{eqnarray}
where $E_C$ is the charging energy, $n$ is the number operator of the Cooper-pairs present on the superconducting island of the transmon, $E_J$ is the Josephson energy and $\varphi$ is the phase difference across the Josephson junction. The driving of the transmon is characterized by $\Omega(t)$. For the description of the impedance the Caldeira-Leggett model \cite{CL} is used, with $H_Z$ being the Hamiltonian for a collection of harmonic oscillators.
The interaction between the bath of oscillators and the transmon is the last term in Eq.~(\ref{eq:lump}), where the bath operator $B$ contains the position degrees of freedom of the oscillators.
The transmon Hamiltonian by itself describes an anharmonic oscillator, and here we restrict the description only to the computational basis of the transmon, namely the two lowest lying levels. In the absence of driving, the truncation process to two-level system yields 
\begin{eqnarray}
H=-\frac{\omega_q}{2}\sigma_z+H_{Z}+\eta e\sigma_yB.
\label{eq:start}
\end{eqnarray}
Equation~(\ref{eq:start}) is the starting point for our description of the transmon qubit as an open quantum system. In Eq.~(\ref{eq:start}), the qubit frequency is $\omega_q=\sqrt{8E_CE_J}-E_C$, while $\sigma_{i}$ are the usual Pauli matrices. The bath operators of the Caldeira-Leggett model are
\begin{eqnarray}
H_Z=\sum_\alpha\omega_\alpha b_\alpha^\dagger b_\alpha,\ B=\sum_\alpha c_\alpha\left(b^\dagger_\alpha+b_\alpha\right),
\label{eq:CL}
\end{eqnarray}
where $c_\alpha$ are coupling strengths with spectral density $J(\omega)=\pi\sum_\alpha|c_\alpha|^2\delta(\omega-\omega_\alpha)$. The circuit analysis of the circuit in Fig.~\ref{fig:circuit} tells us that the spectral density is related to the frequency-dependent impedance by
\begin{eqnarray}
J(\omega)=\textrm{Im}\left(\frac{i\omega Z(\omega)}{1+i\omega\left(C_e+\frac{C_e^2}{C_J+C_e+C_g}\right)Z(\omega)}\right).
\label{eq:SD}
\end{eqnarray}
Here, we assume the oscillators are in thermal equilibrium at inverse temperature $\beta=1/k_B T$, hence the relation between the spectral density and the autocorrelation function is
\begin{eqnarray}
\langle B(t)B(0)\rangle=%\frac{1}{\pi}
\int_0^\infty\textrm{d}\omega\frac{J(\omega)}{\pi}\left(e^{i\omega t}n_\omega+e^{-i\omega t}(1+n_\omega)\right),
\end{eqnarray}
with the Bose function, $n_\omega=(e^{\beta\omega}-1)^{-1}$. 
The harmonic oscillators that constitute the impedance give rise to the quantum noise affecting the transmon. According to the Wiener--Khintchine theorem \cite{Noise2010}, the noise power spectral density is the Fourier transform of the autocorrelation function:
\begin{eqnarray}
S(\omega)=\int_{-\infty}^{\infty}\textrm{d}t\ \langle B(t)B(0)\rangle e^{i\omega t},
\end{eqnarray}
%The bath correlation function, assuming the oscillators are in thermal equilibrium at inverse temperature $\beta$, is given by
%\begin{eqnarray}
%C(t)=\langle B(t)B(0)\rangle=\sum_\alpha\frac{c_\alpha^2}{2m_\alpha\omega_\alpha}\left(n_\alpha e^{i\omega_\alpha t}+(n_\alpha+1)e^{-i\omega_\alpha t}\right),
%\end{eqnarray}
%with $n_\alpha=(e^{\beta\omega_\alpha}-1)^{-1}$ being the Bose function.
%The bath spectral density, which is determined by circuit analysis from the function $K(\omega)$ as in Eq. (8), leads us to the correlation function
thus the relation between the spectral power and spectral density is
\begin{eqnarray}
S_{}(\omega>0)=2J(\omega)(1+n(\omega)),\nonumber\\
S_{}(\omega<0)=2J(-\omega)n(-\omega).
\end{eqnarray}
The quantum nature of the noise manifests itself as the power spectral density is asymmetric: $S(\omega)\neq S(-\omega)$. For convenience, we introduce the symmetrized power spectral density,
\begin{eqnarray}
\bar{S}_{}(\omega)=\frac{1}{2}(S_{}(\omega)+S_{}(-\omega)),
\end{eqnarray}
because it has a direct relation to the spectral density through the fluctuation-dissipation theorem,
\begin{eqnarray}
\bar S_{}(\omega)=\textrm{coth}\left(\frac{\beta\omega}{2}\right)J(\omega).
\end{eqnarray}
We note that, in addition to this symmetrized noise spectral density describing classical noise, the antisymmetric (quantum) part of $S(\omega)$ will also appear via $J(\omega)$.
Finally, the parameter appearing in front of the interaction term is
\begin{eqnarray}
\eta=\frac{2C_e}{C_J+C_g+C_e}\sqrt[4]{\frac{E_J}{4E_C}}.
\label{eq:eta}
\end{eqnarray}
Hence, every necessary component required for the open system descripton of the transmon qubit is supplied to us by circuit theory.

%%%%%NOW THAT WE HAVE THE HAMILTONIAN OF THE SYSTEM-BATH COMPOSITE SYSTEM 
%%%%%WE INTRODUCE THE TCL METHOD AND APPLY IT TO THE OPEN TRANSMON
%%%%%THE SECTION ENDS WITH THE PROPERTIES OF THE DESCRIPTION. NON-MARKOV,CP

\section{\label{sec:TCL}Time-convolutionless projection technique applied}

Before applying the time-convolutionless (TCL) projection operator technique to the superconducting qubit under study, we review the well-known TCL formalism, leading up to Eq.~\eqref{eq:tcl2}.
The theory of open quantum systems aims to describe the dynamics of the reduced, relevant part of a larger composite system. In order to achieve this, one has to trace over the degrees of freedom of the remaining part, which is referred to as the environment. In general, projection operator techniques define the trace over the environment as a formal projection \cite{BRE02}, where the total state of the composite system is projected onto the relevant and irrelevant part. Let $\rho_{\rm tot}$ denote the density operator characterizing the total combined system-environment state. We define the projection superoperator as
$\mathcal P\rho_{\rm tot}=\left(\textrm{Tr}_E\rho_{\rm tot}\right)\otimes\rho_E$,
where Tr$_E$ is the tracing over the environmental degrees of freedom and $\rho_E$ is some fixed state of the environment, called the reference state. 
%The reference state is typically the Gibbs state of the environment at some fixed temperature.
The complementary projection superoperator is defined as $\mathcal Q\rho_{\rm tot}=\rho_{\rm tot}-\mathcal P\rho_{\rm tot}$.
The goal then is to derive a closed equation of motion for $\mathcal P\rho_{\rm tot}$,
immediately leading to a closed equation of motion
for $\rho_S=\textrm{Tr}_E\rho_{\rm tot}$.

The Hamiltonian of a composite system has the general structure
\begin{eqnarray}
H=H_0+\eta H_{\rm int},
%\label{eq:genham}
\end{eqnarray}
with $H_0$ being the free Hamiltonian of the system and environmental degrees of freedom and an interaction term $H_{\rm int}$ between them with dimensionless coupling $\eta$, which will serve later as an expansion parameter. Note that Eq.~(\ref{eq:start}) has exactly this form. Within the interaction picture, the equation of motion for the combined density operator is the von Neumann equation,
\begin{eqnarray}
\frac{\textrm d\rho_I(t)}{\textrm dt}=-i\eta[H_I(t),\rho_I(t)]\equiv\eta\mathcal L(t)\rho_I(t),
\label{eq:vN}
\end{eqnarray}
where the interaction picture operators are related to the Schrödinger picture operators by
\begin{eqnarray}
H_I(t)=e^{iH_0t}H_{\rm int}e^{-iH_0t},\ \rho_I(t)=e^{iH_0t}\rho_{\rm tot}(t)e^{-iH_0t}.
\end{eqnarray}
Earlier, we assumed the reference state $\rho_E$ is the Gibbs state of the environment at some fixed inverse temperature $\beta$. Since the Gibbs state is time independent, applying the projection operators to Eq.~(\ref{eq:vN}) yields closed equations of motion for the projected density operators:
\begin{eqnarray}
\frac{\textrm d}{\textrm dt}\mathcal P\rho_I(t)=\eta\mathcal P\mathcal L(t)\mathcal P\rho_I(t)+\eta\mathcal P\mathcal L(t)\mathcal Q\rho_I(t),\\
\frac{\textrm d}{\textrm dt}\mathcal Q\rho_I(t)=\eta\mathcal Q\mathcal L(t)\mathcal P\rho_I(t)+\eta\mathcal Q\mathcal L(t)\mathcal Q\rho_I(t).
\end{eqnarray}
At this point, the way to proceed is to formally solve the equation of the irrelevant part $\mathcal Q\rho_I(t)$ and substitute it into the equation of the relevant part to obtain the desired closed equation for $\mathcal P\rho_I(t)$. This formal solution can be achieved by two different methods. These yield formally exact, but conceptually different equations of motion for the reduced system density operator \cite{Breuer2004}.

At the end of the first route we would arrive at the famous Nakajima--Zwanzig equation \cite{Nakajima1958,Zwanzig1960}, which is a time-non-local equation containing a memory kernel. Here, we are interested in the second approach \cite{Chaturvedi1979,TCL1}, which yields an equation which does not contain a convolution integral, 
hence it is time-local and generally has the form
\begin{eqnarray}
\frac{\textrm d}{\textrm dt}\mathcal P\rho_I(t)=K(t)\mathcal P\rho_I(t)+I(t)\mathcal Q\rho_I(t_0),
\end{eqnarray}
where the superoperator $K(t)$ is called the TCL generator and $I(t)$ is the inhomogenity term which is only present if there are initial correlations between the reduced system and the environment. The inhomogenity vanishes by assuming that the initial state at the initial time $t_0$ is factorizing, e.g. $\rho_I(t_0)=\rho_S(t_0)\otimes\rho_E$, hence $\mathcal Q\rho_I(t_0)=0$. The TCL generator is obtained by a perturbation expansion \cite{Kubo1963} in to the dimensionless parameter $\eta$,
\begin{eqnarray}
K(t)=\sum_{n=1}^\infty \eta^n K_n(t).
\end{eqnarray}

In our case, the dimensionless coupling is given in Eq.~(\ref{eq:eta}), which can be decreased by increasing the shunting capacitance $C_J$. For the transmon system it is then justified to only consider the first non-vanishing term in the expansion of the TCL generator, namely
\begin{eqnarray}
K(t)=\eta^2\int_{t_0}^t\textrm ds\ \mathcal P\mathcal L(t)\mathcal L(s)\mathcal P+\mathcal O(\eta^4).
\end{eqnarray}
This corresponds to the Born approximation.
The master equation for the open system density operator $\rho_S$ in the interaction picture is
\begin{eqnarray}
\frac{\textrm d}{\textrm dt}\rho_S(t)=-\eta^2\int_{0}^t\textrm ds\  \textrm{Tr}_E\big[H_I(t),[H_I(s),\rho_S(t)\otimes\rho_E]\big],\nonumber\\
\label{eq:tcl2}
\end{eqnarray}
where without loss of generality, we defined the initial time $t_0=0$.

We now return to our analysis of the non-Markovian dynamics of the superconducting qubit.
By considering the Hamiltonian of the open transmon system in Eq.~(\ref{eq:start}), the TCL-Born master equation of Eq.~(\ref{eq:tcl2}) in the Schrödinger picture becomes
\begin{multline}
\dot\rho(t)= 
-i[H(t),\rho(t)]\\
+\eta^2\sum_{k,l=\pm}d_{kl}(t)\left(\sigma_k\rho(t)\sigma_l^\dagger-\frac{1}{2}\{\sigma_l^\dagger\sigma_k,\rho(t)\}\right).
\label{eq:tcl-born}
\end{multline}
The unitary part of Eq.~(\ref{eq:tcl-born}) consists of the qubit Hamiltonian, which acquires a time-dependent Lamb shift due to interactions with the environment,
\begin{eqnarray}
H(t)&=&-\frac{1}{2}\left(\omega_q+\eta^2\omega_{\rm LS}(t)\right)\sigma_z,\\
\omega_{\rm LS}(t)&=&\frac{2e^2}{\pi}\int_0^t\textrm d\tau\sin\omega_q\tau\int_0^\infty\textrm d\omega\ \bar S_{}(\omega)\cos\omega\tau.\nonumber
\label{eq:LS}
\end{eqnarray}
The second part of Eq.~(\ref{eq:tcl-born}) is the dissipator part, which describes the relaxation of the transmon caused by its environment through time-dependant rates,
\begin{widetext}
\begin{eqnarray}
d_{kl}(t)=\begin{pmatrix}
\gamma_+(t)&-\frac{\gamma_+(t)+\gamma_-(t)}{2}-i\omega_{\rm LS}(t)\\
-\frac{\gamma_+(t)+\gamma_-(t)}{2}+i\omega_{\rm LS}(t)&\gamma_-(t)
\end{pmatrix},
\label{eq:d}
\end{eqnarray}
\begin{eqnarray}
\gamma_{\pm}(t)=\frac{2e^2}{\pi}\int_0^t\textrm d\tau\cos\omega_q\tau\int_0^\infty\textrm d\omega\ \bar S_{}(\omega)\cos\omega\tau\pm\frac{2e^2}{\pi}\int_0^t\textrm d\tau\sin\omega_q\tau\int_0^\infty\textrm d\omega\ J(\omega)\sin\omega\tau.
\label{eq:rates}
\end{eqnarray}
\end{widetext}
%The off-diagonal elements of the decoherence matrix $d_{kl}$ correspond to non-secular terms in the master equation. In the following, we show how the presence of these terms have striking consequences on the non-Markovian nature of the equation.

\subsection{\label{sec:TCLA}Eternal non-Markovianity}
\begin{figure}[t]
\includegraphics[scale=.8]{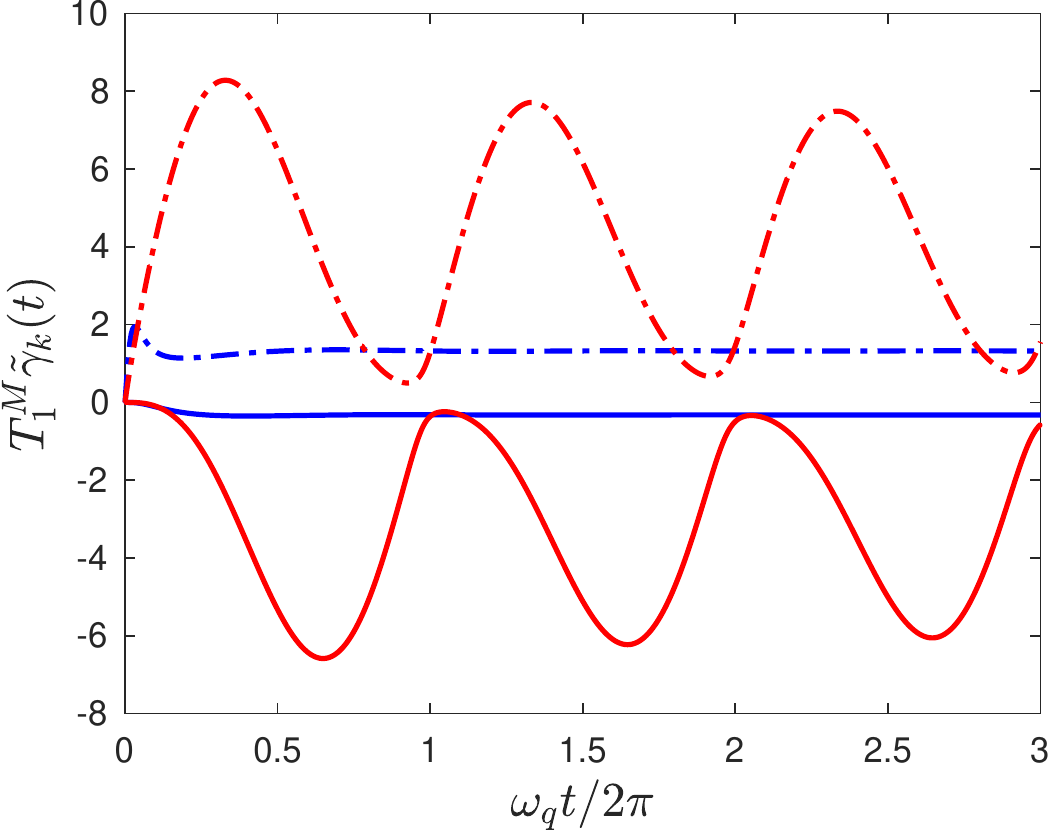}% Here is how to import EPS art
\caption{\label{fig:rates}Dimensionless canonical rates $\tilde\gamma_1(t)$ (dash-dotted lines) and $\tilde\gamma_2(t)$ (solid lines) appearing in Eq.~(\ref{eq:tcl-born}) for an environment with Ohmic (blue) and $1/f^\alpha$ (red) spectral density. The parameters for this plot are the following: in the Ohmic case $\omega_c=5\omega_q$, $e^2R\eta^2e^{-\omega_q/\omega_c}=10^{-4}$, and in the $1/f^\alpha$ case $Ae^2\eta^2/\omega_q^{\alpha+1}=10^{-4}$, $\alpha=0.95$. These parameters correspond to equal Markovian relaxation times $T_1^M$ for the two spectral densities \eqref{eq:ohm} and \eqref{eq:alfa}.}
\end{figure}

The master equation for the open transmon system in Eq.~(\ref{eq:tcl-born}) is in a generalized Lindblad form with time-dependent rates. It is possible to rewrite the entire master equation in the canonical Lindblad form through the diagonalization of the decoherence matrix $d_{kl}$,
\begin{multline}
\dot\rho(t)=-i[H(t),\rho(t)]\\
+\eta^2\sum_{k=1,2}\tilde\gamma_k(t)\left(L_k(t)\rho(t)L_k^\dagger(t)-\frac{1}{2}\{L_k^\dagger(t)L_k(t),\rho(t)\}\right).
\end{multline}
Here, the canonical decay rates $\tilde\gamma_k$ are the eigenvalues of $d_{kl}$ and are given by
\begin{multline}
\tilde\gamma_{1,2}(t)=\frac{\gamma_+(t)+\gamma_-(t)}{2}\\
\pm\sqrt{\frac{(\gamma_+(t)+\gamma_-(t))^2}{4}+\frac{(\gamma_+(t)-\gamma_-(t))^2}{4}+\omega_{\rm LS}(t)^2}.
\label{eq:tildeg}
\end{multline}
According to the definition mentioned in the Introduction, a Markovian process is CP-divisible at all intermediate times during evolution. Any master equation in the canonical Lindblad form describes a Markovian process whenever the rates $\tilde\gamma_k(t)\geq0$, for all $k$ and $t$. However, it is easy to see that in our case $\tilde\gamma_2(t)\le 0$ at all times $t>0$, see also Fig.~\ref{fig:rates} for an example. In light of this observation, we conclude that the idle time evolution of the open transmon system is eternally non-Markovian \cite{Megier2017}%, unless $\tilde\gamma_2(t)\equiv 0$
. The eternally non-Markovian dynamics reduces to simple Markovian dynamics if $\tilde\gamma_2(t)\equiv 0$, which according to Eq.~\eqref{eq:tildeg} requires $\gamma_+(t)=\gamma_-(t)$ and $\omega_{\rm LS}(t)=0$ for all times. This situation occurs if the autocorrelation function of the bath operators is $\langle B(t)B(0)\rangle\sim\delta(t)$, which corresponds to a trivially memoryless bath.
We note that at this point we have not specifed the actual temperature of the bath nor the details of the qubit, hence we find that the eternal non-Markovianity for the evolution process appears generally in the case of transverse qubit-bath coupling.
%
%According to the GKSL theorem, a master equation in canonical Lindblad form describes Markovian evolution if the canonical rates are positive at all times. If at any time one of the rates $\tilde\gamma_k$ is negative, then through the corresponding decoherence channel $L_k$ the system may recohere, reversing earlier decay processes. 

\subsection{\label{sec:TCLB}Solution and complete positivity}

The master equation for the open transmon system preserves the Hermiticity and the trace of the density operator. Through the solution of the master equation, we can construct the dynamical map that transforms the initial state of the qubit $\rho(0)$ to the state at a later time $\rho(t)$. 
Let the computational basis be $|j\rangle$, $j=0,1$. We define the index $a$ as the ordered pairs $(j_1,j_2)$, then the dynamical map can be expressed by the matrices $\tau_a=|j_1\rangle\langle j_2|$ as \cite{Andersson2007}
\begin{eqnarray}
\rho(t)=\sum_{a,b}C_{ab}(t)\tau_a\rho(0)\tau_b^\dagger.
\label{eq:map}
\end{eqnarray}
The solution of the differential equation system in Eq.~(\ref{eq:tcl-born}) is expressed within the Choi matrix \cite{Choi1975},
\begin{widetext}
\begin{eqnarray}
C_{ab}(t)=\begin{pmatrix}
\frac{1}{2}\left(1+Z(t)+e^{-\Gamma(t)}\right)&0&0&x_+(t)e^{i\varphi(t)-\frac{\Gamma(t)}{2}}\\
0&\frac{1}{2}\left(1+Z(t)-e^{-\Gamma(t)}\right)&x_-(t)e^{i\varphi(t)-\frac{\Gamma(t)}{2}}&0\\
0&x_-^*(t)e^{-i\varphi(t)-\frac{\Gamma(t)}{2}}&\frac{1}{2}\left(1-Z(t)-e^{-\Gamma(t)}\right)&0\\
x_+^*(t)e^{-i\varphi(t)-\frac{\Gamma(t)}{2}}&0&0&\frac{1}{2}\left(1-Z(t)+e^{-\Gamma(t)}\right)
\end{pmatrix}.
\end{eqnarray}
\end{widetext}
%In the computational basis $|j\rangle$, $j=0,1$, the matrix elements of the density operator are $\langle i|\rho|j\rangle$. The solution of the differential equation system in Eq.~(\ref{eq:tcl-born}) is expressed as
%\begin{eqnarray}
%\rho_{00}(t)=\frac{1}{2}\left(1+Z(t)+e^{-\Gamma(t)}(2\rho_{00}(0)-1)\right),\\
%\rho_{11}(t)=\frac{1}{2}\left(1-Z(t)+e^{-\Gamma(t)}(2\rho_{11}(0)-1)\right),\\
%\rho_{01}(t)=e^{i\varphi(t)-\frac{\Gamma(t)}{2}}\left(x(t)\rho_{01}(0)+y(t)\rho_{10}(0)\right),\\
%\rho_{10}(t)=e^{-i\varphi(t)-\frac{\Gamma(t)}{2}}\left(x^*(t)\rho_{10}(0)+y^*(t)\rho_{01}(0)\right).
%\end{eqnarray}
Here, the decay function is
\begin{eqnarray}
\Gamma(t)=\eta^2\int_0^t \textrm ds\ \left(\gamma_+(s)+\gamma_-(s)\right),
\label{eq:decay}
\end{eqnarray}
while the relaxation function, to which the diagonal elements relax, is
\begin{eqnarray}
Z(t)=\eta^2e^{-\Gamma(t)}\int_0^t \textrm ds\ e^{\Gamma(s)}\left(\gamma_+(s)-\gamma_-(s)\right).
\end{eqnarray}
The phase function of the off-diagonal elements is
\begin{eqnarray}
\varphi(t)=\omega_qt+\eta^2\int_0^t\textrm ds\ \omega_{\rm LS}(s),
\end{eqnarray}
and the coherence functions are determined by the following system of differential equations \cite{RWAnm},
\begin{eqnarray}
\dot x_\pm=\eta^2\left(-i\omega_{LS}(t)-\frac{\gamma_+(t)+\gamma_-(t)}{2}\right)e^{-2i\varphi(t)}x_\mp^*,\label{eq:x}
%\dot y=\eta^2\left(-i\omega_{\rm LS}(t)-\frac{\gamma_+(t)+\gamma_-(t)}{2}\right)e^{-2i\varphi(t)}x^*,
%\label{eq:y}
\end{eqnarray}
with initial conditions $x_+(0)=1$, $x_-(0)=0$. 

The Choi matrix (\ref{eq:map}) makes it easy to analyze the complete positivity property of the dynamical map. A map is completely positive whenever it admits a Kraus operator sum representation \cite{Kraus1983}. We can only transform Eq.~(\ref{eq:map}) into the Kraus representation if the Choi matrix is positive semidefinite, $C_{ab}\geq0$, hence we are able to construct the conditions for complete positivity from the eigenvalues of the Choi matrix,
\begin{eqnarray}
\lambda_{1,2}=\frac{1}{2}\left(1+e^{-\Gamma(t)}\pm\sqrt{Z(t)^2+4e^{-\Gamma(t)}|x_+(t)|^2}\right),\\
\lambda_{3,4}=\frac{1}{2}\left(1-e^{-\Gamma(t)}\pm\sqrt{Z(t)^2+4e^{-\Gamma(t)}|x_-(t)|^2}\right).
\end{eqnarray}
It is easy to see that $\lambda_1\ge\max(0,\lambda_2)$ and $\lambda_3\ge\lambda_4$, thus complete positivity requires
\begin{eqnarray}
1\pm e^{-\Gamma(t)}\geq\sqrt{Z(t)^2+4e^{-\Gamma(t)}|x_\pm(t)|^2}.
\label{eq:cond1}
%1-e^{-\Gamma(t)}\geq\sqrt{Z(t)^2+4e^{-\Gamma(t)}|y(t)|^2}.
%\label{eq:cond2}
\end{eqnarray}
We note that Eq.~(\ref{eq:x}) implies $|x_+(t)|^2-|x_-(t)|^2=1$, thus the conditions in Eq.~(\ref{eq:cond1}) are equivalent.
The \textit{necessary} condition is
\begin{eqnarray}
\Gamma(t)\geq0,
\end{eqnarray}
because $\Gamma(t)$ being negative for some time implies~$1-e^{-\Gamma(t)}<0$ and Eq.~(\ref{eq:cond1}) cannot be satisfied. With the definition
\begin{eqnarray}
f_\pm(t)\equiv\int_0^t \textrm ds\ e^{\Gamma(s)}\gamma_\pm(s),
\end{eqnarray}
the \textit{sufficient} condition for complete positivity becomes
\begin{eqnarray}
e^{-\Gamma(t)}f_+(t)f_-(t)\geq \frac{|x_-(t)|^2}{\eta^4}.
\label{eq:suff}
\end{eqnarray}
The necessary and sufficient conditions enable us to numerically check the complete positivity of the dynamical map in Eq.~(\ref{eq:map}). Here, we confirm the dynamical map is CP for any values of the parameters used in this paper.

\section{\label{sec:examples}Results at zero temperature}

Within this section, we analyze the non-Markovian dynamics of the qubit for two distinct types of bath: the Ohmic and the $1/f^\alpha$ bath, both at $T=0$. The spectral density of the Ohmic bath is
\begin{eqnarray}
J_O(\omega)=R\omega e^{-\frac{\omega}{\omega_c}},
\label{eq:ohm}
\end{eqnarray}
with $\omega_c$ being a high frequency cutoff. The spectral density of the $1/f^\alpha$ bath is
\begin{eqnarray}
J_\alpha(\omega)=\frac{A}{\omega^\alpha}.
\label{eq:alfa}
\end{eqnarray}

We also wish to compare our results with Markovian predictions in order to identify the major consequences of non-Markovianity. 
The Markovian limit corresponds to the extension of the time integrals to infinity in Eq.~\eqref{eq:LS} and Eq.~\eqref{eq:rates}, which corresponds to the replacement of the time-dependent rates and time-dependent Lamb shift with their asymptotic values. These are
\begin{eqnarray}
\gamma_\pm^M=e^2\left(\bar S_{}(\omega_q)\pm J(\omega_q)\right),\nonumber\\
\omega^M_{\rm LS}=\omega_q\frac{2e^2}{\pi}\ P \int_0^\infty \mathrm d\omega\ \frac{\bar S_{}(\omega)}{\omega_q^2-\omega^2},
\label{eq:markov1}
\end{eqnarray}
where $P$ denotes principal value integration. At zero temperature $\bar S_{}(\omega)= J(\omega)$, which implies $\gamma_-^M=0$. Thus, according to the sufficient condition in Eq.~\eqref{eq:suff}, the Markovian approximation invalidates the complete positivity of the dynamical map at $T=0$. In order to ensure complete positivity of the dynamical map, the Markov approximation has to be supplemented with the secular approxmiation, consequently Eq.~\eqref{eq:tcl-born} reduces to the well-known Lindblad equation, with rates and Lamb shift given by Eq.~\eqref{eq:markov1}. Hereafter, we refer to the Markovian solution as the solution of this Lindblad equation.

\subsection{Qubit precession}
The discussion of non-Markovian effects on the qubit is most conveniently done through the behaviour of the expectation value of $\sigma_x$ from the dynamical map Eq.~\eqref{eq:map},
\begin{eqnarray}
\langle\sigma_x(t)\rangle=2e^{-\frac{\Gamma(t)}{2}}\textrm{Re}\left(e^{i\varphi(t)}(\rho_{01}(0)x_+(t)+\rho_{10}(0)x_-(t))\right).\nonumber\\
\label{eq:SX}
\end{eqnarray}
The first step in the evaluation of $\langle\sigma_x(t)\rangle$ consists in a calculation of the time-dependent rates and Lamb shift from Eq.~\eqref{eq:LS} and Eq.~\eqref{eq:rates} for a given bath spectral density.
Secondly, we need to solve the differential equation system Eq.~\eqref{eq:x}. For the Markovian case, the secular approximation implies $x_+(t)\equiv1$ and $x_-(t)\equiv0$. 
\begin{figure}[t]
\includegraphics[scale=.8]{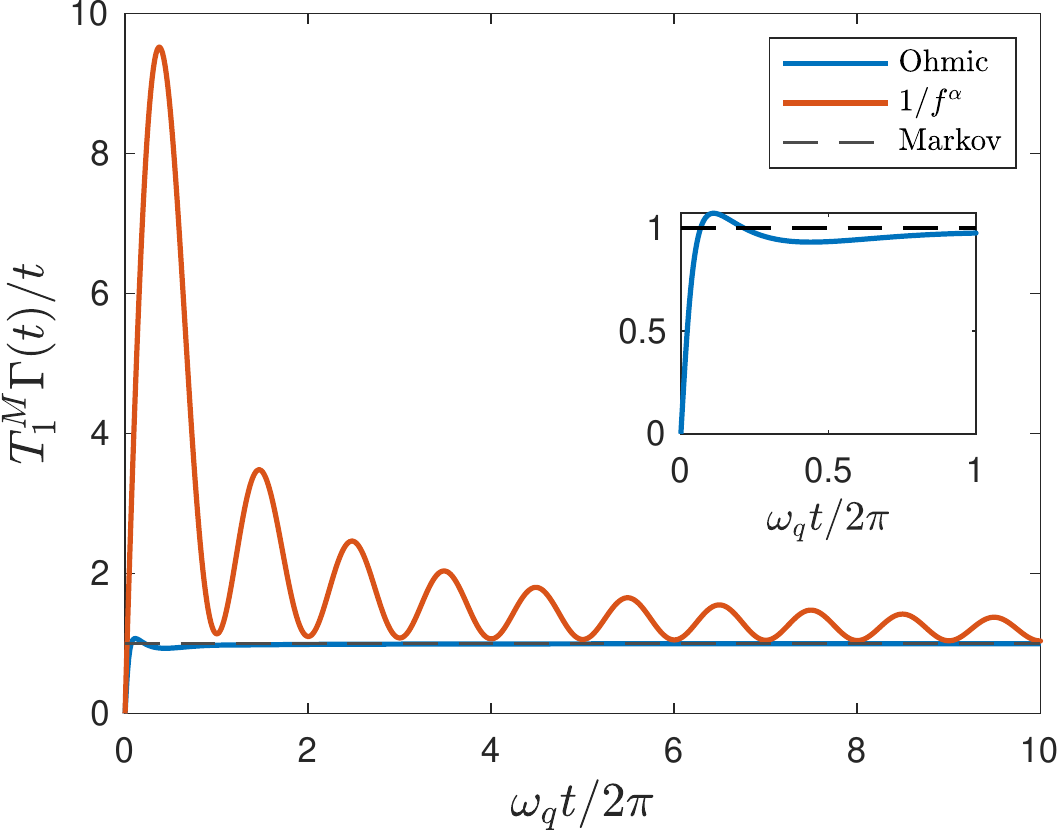}% Here is how to import EPS art
\caption{\label{fig:gamma}Decay function $\Gamma(t)$ divided by time in units of the Markovian decay rate $1/T^M_1=2/T^M_2$. The blue and red curves correspond to the non-Markovian dynamics in the presence of an Ohmic and $1/f^\alpha$ bath, respectively. The dashed line denotes the Markovian solution. The inset shows a magnification for short times in the Ohmic case. The parameters for this plot are $\omega_c=3\omega_q$, $\alpha=0.95$.}
\end{figure}

In the noiseless case $\eta=0$, Eq.~\eqref{eq:SX} describes the precession of the qubit state  with frequency $\omega_q$ around the $Z$-axis of the Bloch sphere. In the presence of noise, the Markovian solution introduces the well-known exponential damping with the timescale of this decay being 
$T_2=\frac{2}{\eta^2(\gamma^M_++\gamma_-^M)}$, 
and it also shifts the precession frequency by $\eta^2\omega_{\rm LS}^M$. 
\begin{figure*}[t]
\includegraphics[scale=.8]{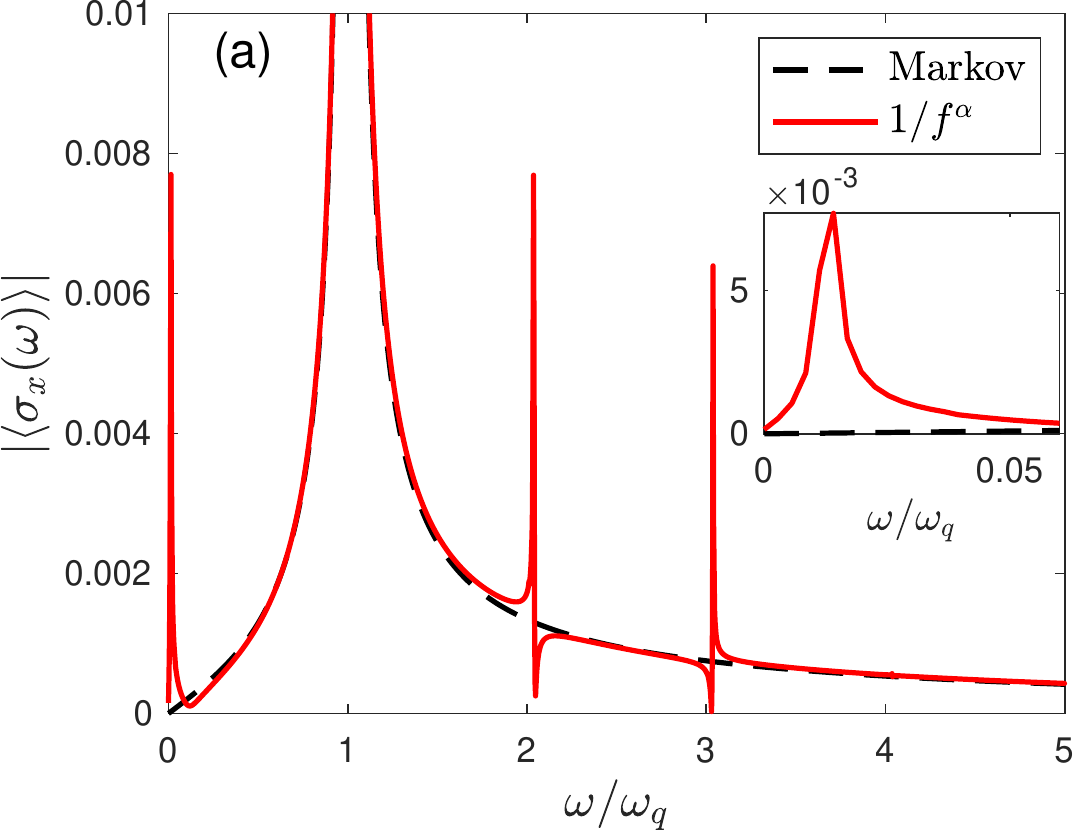}\quad \includegraphics[scale=.8]{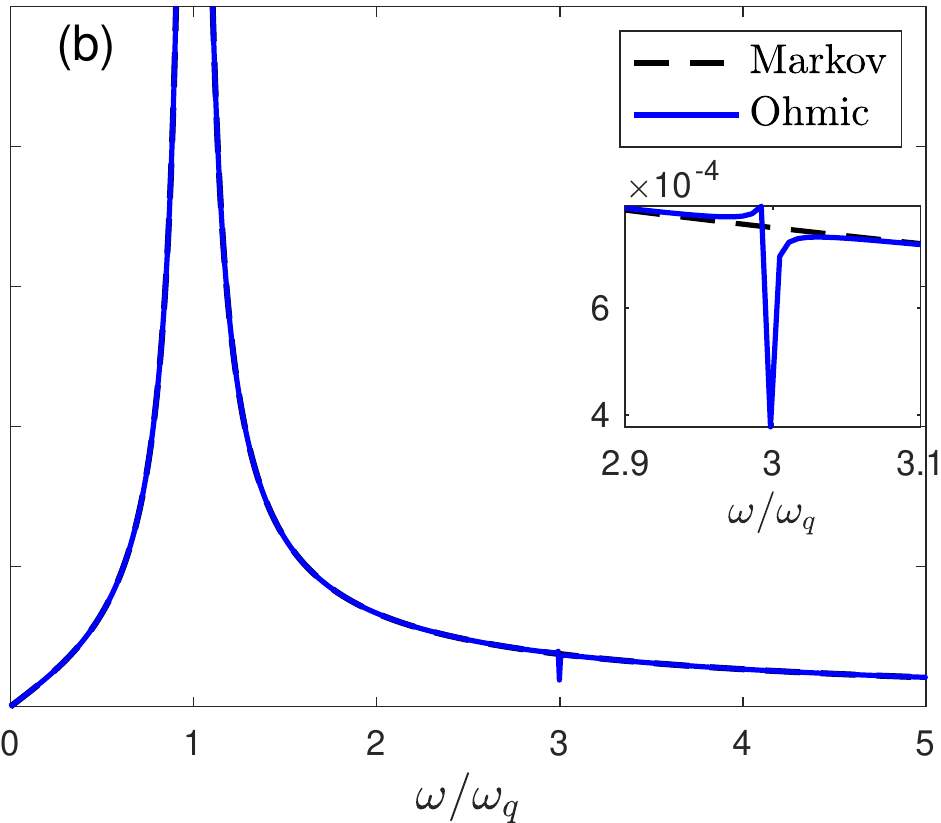} \caption{\label{fig:fft}The absolute value of the Fourier transform of $\langle\sigma_x(t)\rangle$ for the (a) $1/f^\alpha$ and (b) Ohmic bath spectra. The blue and red lines correspond to the Markovian and non-Markovian solution, respectively. The Inset of (a) shows the low frequency behaviour. The Inset of (b) shows a detail around $\omega=3\omega_q$. For both cases, we chose the initial state $\rho_{01}(0)=\rho_{10}(0)=1/2$. The parameters used for (a) are $Ae^2\eta^2/\omega_q^{\alpha+1}=10^{-3}$ (this value corresponds to $T_2\sim \mu s$), $\alpha=0.95$. The parameters used for (b) are $e^2 R\eta^2e^{-\omega_q/\omega_c}=10^{-3}$ (this value corresponds to $T_2\sim \mu s$), $\omega_c=5\omega_q$.}
\end{figure*}

Non-Markovianity changes this picture in two ways. First, the damping becomes non-exponential. In Fig.~\ref{fig:gamma}, we depict the decay function from Eq.~\eqref{eq:decay} divided by time, which in the Markov case equals $1/T_2$. From Fig.~\ref{fig:gamma}, it is evident that the Ohmic noise only slightly deviates from the Markovian approximation and that the Ohmic decay function assumes an exponential form after a short time. On the other hand, the $1/f^\alpha$ noise approaches the Markovian behaviour on a much larger timescale. Additionally, the oscillations present in the decay function give rise to periodic recoherence, a purely non-Markovian phenomenon. Instead of continuously losing information to the environment, at certain times the open system regains some of its lost information, which is possible due to the channels that have negative decay rates.

The second and probably more important effect of non-Markovianity is the introduction of additional frequencies to the qubit precession. Through the non-secular terms in the TCL-Born master equation, higher harmonics enter the solution of the differential equation system Eq.~\eqref{eq:x} which cause the appearance of additional frequencies far from $\omega_q$ in the precession, see Fig.~\ref{fig:fft}. The presence of these frequencies has important consequences on the free qubit evolution, which can be revealed by applying single qubit gates, as discussed in the next subsection.

\subsection{Revealing non-Markovian effects using single qubit gates}

Single qubit gates, in other words operations that rotate the qubit state around the Bloch sphere in a controlled fashion, are designed for transmons by driving the qubit on resonance with the qubit frequency $\omega_q$. Seen from the frame which rotates with the qubit frequency, an in-phase drive corresponds to rotation around the $X$-axis, while a pulse with a relative phase of $\pi/2$ causes rotation around the $Y$-axis \cite{Krantz2019}. 
%This constitutes as the $XY$ control of a transmon. Here, it is cruical to understand, if the drive is off-resonant with the qubit frequency then it results in undesired rotational effects on the qubit state.

Non-Markovian noise introduces extra frequencies far from $\omega_q$ to the precession of the qubit, hence in the rotating frame a portion of the qubit state is not stationary. When a gate is in operation, these additional terms rotate undesirably and cause gate errors that are not taken into account in a Markovian description. These errors do not only depend on the timing of the gate, i.e., \emph{when} we rotate the state, but also on the rotation axis. To showcase this, consider the following scenario, akin to a Ramsey experiment. We prepare the qubit in its $|0\rangle$ state and rotate it by $\pi/2$ around the $Y$-axis of the Bloch sphere, thus producing the superposition state $|0\rangle+|1\rangle$. After a delay time, we rotate the state back to the north pole by a $-\pi/2$ rotation around the $Y$-axis and measure the state. We replicate this experiment, but this time we rotate the qubit state around the $X$-axis of the Bloch sphere first by $-\pi/2$ producing  $|0\rangle+i |1\rangle$, and after the same delay time back around the $X$-axis by $\pi/2$. 

\begin{figure}[t]
\includegraphics[scale=.75]{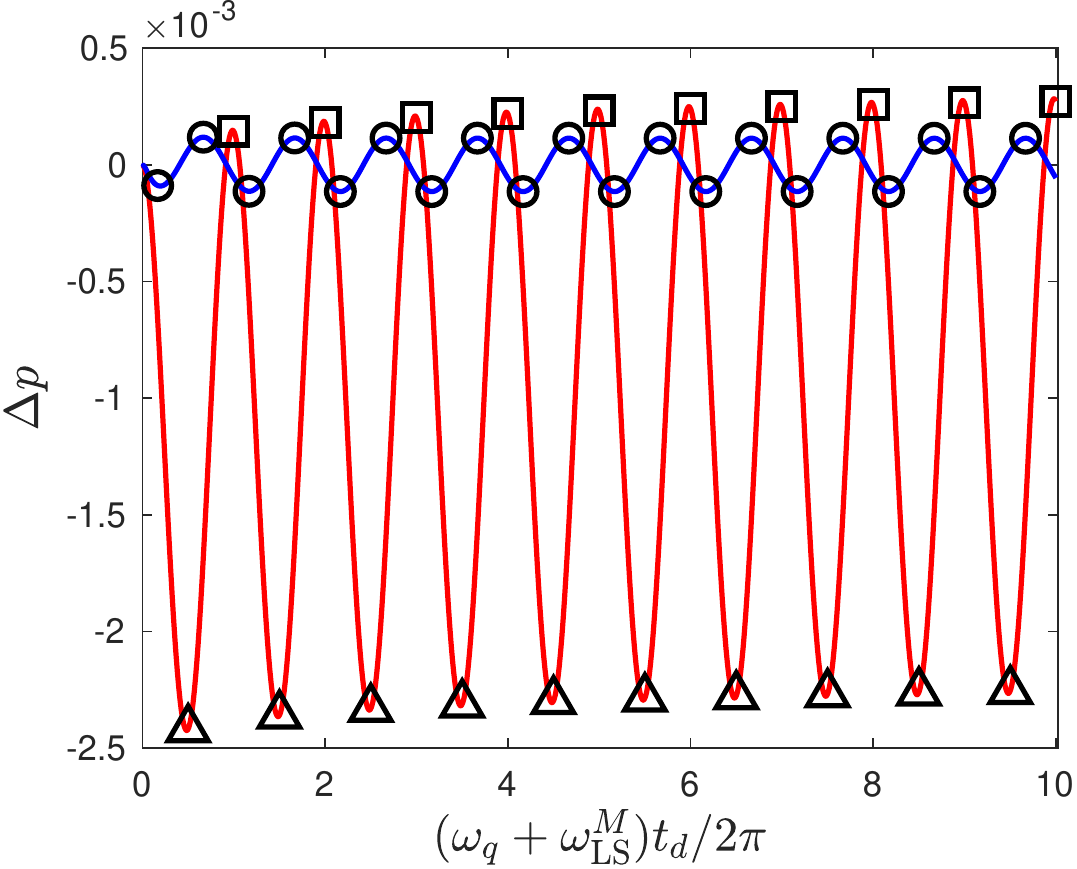}% Here is how to import EPS art
\caption{\label{fig:prob}Smoking-gun signature of non-Markovian dynamics. The difference of probabilities $\Delta p$ of measuring the ground state after rotations around the $X$ and $Y$-axis of the Bloch sphere in the presence of $1/f^\alpha$ noise (red line) and Ohmic noise (blue line).
The horizontal axis shows the delay time in units of a whole precession period. In the Markovian case one finds $\Delta p=0$. For the $1/f^\alpha$ noise the difference of the probabilities is much lower when the delay time allows a full precession (black squares), because in that case the higher harmonics are aligned with the main frequency and only the slowly precessing term causes a difference. $|\Delta p|$ is largest when the delay time is half a period (black triangles).
While for $1/f^\alpha$ this effect is of the order $\Delta p\approx 10^{-3}$, it is much smaller ($\approx 10^{-5}$) for Ohmic noise, for which the maxima appear at one-sixth and two-thirds of a period (black circles).
The parameters used for this plot are $Ae^2\eta^2/\omega_q^{\alpha+1}=10^{-4}$, $e^2 R\eta^2e^{-\omega_q/\omega_c}=10^{-4}$ corresponding to $T_2\sim 10\,\mu s$, and $\alpha=0.95$, $\omega_c=3\omega_q$.}
\end{figure}

In the absence of noise, the probability of measuring $|0\rangle$ is equal at all times in both cases. Considering the noise, by using the dynamical map in Eq.~\eqref{eq:map}, we calculate the difference of these probabilites as a function of the delay time $t_d$,
\begin{multline}
\Delta p=  p(|0\rangle|YY)- p(|0\rangle|XX)\\
=e^{-\frac{\Gamma(t_d)}{2}}\left(\cos\varphi(t_d)\textrm{Re}(x_-(t_d))-\sin\varphi(t_d)\textrm{Im}(x_-(t_d))\right).
\label{eq:prob}
\end{multline}
The probabilities differ due to the appearance of $x_-(t)$, which is completely disregarded in the Markovian picture. As a result, the Markovian treatment of the noise predicts equal probabilities $\Delta p=0$ for arbitrary delay times. The function $x_-(t)$, which originates from the non-secular terms, introduces the extra frequencies appearing in the qubit precession which highlights the difference between Markovian and non-Markovian noise. Depending on the delay time, parts of the qubit state corresponding to the additional precession frequencies are not alligned with the main portion of the state which precesses with the qubit frequency, see Fig.~\ref{fig:prob}. After the state is rotated, the non-alligned parts still linger in the system as long as the relaxation time allows and they effectively act as the memory of the state prior to the gate operation. The presence of these remnants immediately lowers the return probability compared to the Markovian estimate. Furthermore, the idle evolution of the qubit during the delay time is sensitive to its initial state which explains the difference between the probabilities depending on the rotation axis. We remark that we assumed instantaneous and perfect rotations in the calculation of the probability difference in Eq.~\eqref{eq:prob}. Gate imperfections could further enhance the difference.

The physical background of the differing probabilities in Eq.~\eqref{eq:prob} is a matter of a symmetry property which differ in the Markovian and non-Markovian cases. 
The above described experimental setups are transformed into each other by a $\pi/2$ rotation around the $Z$-axis of the Bloch sphere, this transformation is described by $R_z=e^{-\frac{i\pi}{4}\sigma_z}$. 
The system-bath Hamiltonian in Eq.~\eqref{eq:start} is not invariant under this symmetry, and the lack of this symmetry is carried over to the non-Markovian master equation. The secular approximation in the Markovian picture introduces the symmetry described by $R_z$ artificially. Indeed, since $R_z\sigma_\pm R_z^\dagger=\mp i\sigma_\pm$, the transformation leaves the secular terms in Eq.~\eqref{eq:tcl-born} invariant and it adds a minus sign to the non-secular terms. Consequently, the symmetry is present in the Markovian case, however it is absent in the non-Markovian description and it leads to differing probabilities. This suggests the presented result of Eq.~\eqref{eq:prob} depends on the form of the system-bath interaction Hamiltonian. 

\section{\label{sec:conclusion}Conclusion}

We have studied the non-Markovian effects of an electromagnetic environment that influences a superconducting transmon qubit. The composite qubit-environment Hamiltonian is obtained by applying quantum circuit theory to the lumped circuit model describing the transmon. Then, we used the time-convolutionless projection technique to estabish a non-Markovian master equation describing the reduced dynamics of the qubit. 

We found two major non-Markovian effects on the idle evolution of the open transmon system. First, the qubit decay is non-exponential and it allows periodic recoherence which is possible through the channel with negative decay rate. Secondly, the spectral analysis of the decaying qubit precession reveals the presence of additional frequencies far from the qubit frequency which are not contained in the traditional Markovian theory. These novel precession terms should be observable in Ramsey experiments. Furthermore, we have shown how in a Ramsey experiment, non-Markovian noise causes an imbalance between probabilities corresponding to different rotation axes, i.e. the outcomes should depend on whether $X$ or $Y$ pulses are applied.  As a result, for transmons the smoking gun of non-Markovian noise could be found by comparing the results of Ramsey measurements for $X$ and $Y$ pulses.

\begin{acknowledgments}
We acknowledge the support from the German Ministry for Education and Research, under the QSolid project, Grant No.~13N16167.

\end{acknowledgments}

\bibliography{scnonmarkov}% Produces the bibliography via BibTeX.

\begin{thebibliography}{60}%
\makeatletter
\providecommand \@ifxundefined [1]{%
 \@ifx{#1\undefined}
}%
\providecommand \@ifnum [1]{%
 \ifnum #1\expandafter \@firstoftwo
 \else \expandafter \@secondoftwo
 \fi
}%
\providecommand \@ifx [1]{%
 \ifx #1\expandafter \@firstoftwo
 \else \expandafter \@secondoftwo
 \fi
}%
\providecommand \natexlab [1]{#1}%
\providecommand \enquote  [1]{``#1''}%
\providecommand \bibnamefont  [1]{#1}%
\providecommand \bibfnamefont [1]{#1}%
\providecommand \citenamefont [1]{#1}%
\providecommand \href@noop [0]{\@secondoftwo}%
\providecommand \href [0]{\begingroup \@sanitize@url \@href}%
\providecommand \@href[1]{\@@startlink{#1}\@@href}%
\providecommand \@@href[1]{\endgroup#1\@@endlink}%
\providecommand \@sanitize@url [0]{\catcode `\\12\catcode `\$12\catcode
  `\&12\catcode `\#12\catcode `\^12\catcode `\_12\catcode `\%12\relax}%
\providecommand \@@startlink[1]{}%
\providecommand \@@endlink[0]{}%
\providecommand \url  [0]{\begingroup\@sanitize@url \@url }%
\providecommand \@url [1]{\endgroup\@href {#1}{\urlprefix }}%
\providecommand \urlprefix  [0]{URL }%
\providecommand \Eprint [0]{\href }%
\providecommand \doibase [0]{https://doi.org/}%
\providecommand \selectlanguage [0]{\@gobble}%
\providecommand \bibinfo  [0]{\@secondoftwo}%
\providecommand \bibfield  [0]{\@secondoftwo}%
\providecommand \translation [1]{[#1]}%
\providecommand \BibitemOpen [0]{}%
\providecommand \bibitemStop [0]{}%
\providecommand \bibitemNoStop [0]{.\EOS\space}%
\providecommand \EOS [0]{\spacefactor3000\relax}%
\providecommand \BibitemShut  [1]{\csname bibitem#1\endcsname}%
\let\auto@bib@innerbib\@empty
%</preamble>
\bibitem [{\citenamefont {Gambetta}\ \emph {et~al.}(2017)\citenamefont
  {Gambetta}, \citenamefont {Chow},\ and\ \citenamefont
  {Steffen}}]{Gambetta2017}%
  \BibitemOpen
  \bibfield  {author} {\bibinfo {author} {\bibfnamefont {J.~M.}\ \bibnamefont
  {Gambetta}}, \bibinfo {author} {\bibfnamefont {J.~M.}\ \bibnamefont {Chow}},\
  and\ \bibinfo {author} {\bibfnamefont {M.}~\bibnamefont {Steffen}},\ }\href
  {https://doi.org/10.1038/s41534-016-0004-0} {\bibfield  {journal} {\bibinfo
  {journal} {npj Quantum Information}\ }\textbf {\bibinfo {volume} {3}},\
  \bibinfo {pages} {2} (\bibinfo {year} {2017})}\BibitemShut {NoStop}%
\bibitem [{\citenamefont {Wendin}(2017)}]{Wendin2017}%
  \BibitemOpen
  \bibfield  {author} {\bibinfo {author} {\bibfnamefont {G.}~\bibnamefont
  {Wendin}},\ }\href {https://doi.org/10.1088/1361-6633/aa7e1a} {\bibfield
  {journal} {\bibinfo  {journal} {Reports on Progress in Physics}\ }\textbf
  {\bibinfo {volume} {80}},\ \bibinfo {pages} {106001} (\bibinfo {year}
  {2017})}\BibitemShut {NoStop}%
\bibitem [{\citenamefont {Martinis}\ \emph {et~al.}(2020)\citenamefont
  {Martinis}, \citenamefont {Devoret},\ and\ \citenamefont
  {Clarke}}]{Martinis2020}%
  \BibitemOpen
  \bibfield  {author} {\bibinfo {author} {\bibfnamefont {J.~M.}\ \bibnamefont
  {Martinis}}, \bibinfo {author} {\bibfnamefont {M.~H.}\ \bibnamefont
  {Devoret}},\ and\ \bibinfo {author} {\bibfnamefont {J.}~\bibnamefont
  {Clarke}},\ }\href {https://doi.org/10.1038/s41567-020-0829-5} {\bibfield
  {journal} {\bibinfo  {journal} {Nature Physics}\ }\textbf {\bibinfo {volume}
  {16}},\ \bibinfo {pages} {234} (\bibinfo {year} {2020})}\BibitemShut
  {NoStop}%
\bibitem [{\citenamefont {Oliver}\ and\ \citenamefont
  {Welander}(2013)}]{scqubit1}%
  \BibitemOpen
  \bibfield  {author} {\bibinfo {author} {\bibfnamefont {W.~D.}\ \bibnamefont
  {Oliver}}\ and\ \bibinfo {author} {\bibfnamefont {P.~B.}\ \bibnamefont
  {Welander}},\ }\href {https://doi.org/10.1557/mrs.2013.229} {\bibfield
  {journal} {\bibinfo  {journal} {MRS Bulletin}\ }\textbf {\bibinfo {volume}
  {38}},\ \bibinfo {pages} {816–825} (\bibinfo {year} {2013})}\BibitemShut
  {NoStop}%
\bibitem [{\citenamefont {Koch}\ \emph {et~al.}(2007)\citenamefont {Koch},
  \citenamefont {Yu}, \citenamefont {Gambetta}, \citenamefont {Houck},
  \citenamefont {Schuster}, \citenamefont {Majer}, \citenamefont {Blais},
  \citenamefont {Devoret}, \citenamefont {Girvin},\ and\ \citenamefont
  {Schoelkopf}}]{Transmon1}%
  \BibitemOpen
  \bibfield  {author} {\bibinfo {author} {\bibfnamefont {J.}~\bibnamefont
  {Koch}}, \bibinfo {author} {\bibfnamefont {T.~M.}\ \bibnamefont {Yu}},
  \bibinfo {author} {\bibfnamefont {J.}~\bibnamefont {Gambetta}}, \bibinfo
  {author} {\bibfnamefont {A.~A.}\ \bibnamefont {Houck}}, \bibinfo {author}
  {\bibfnamefont {D.~I.}\ \bibnamefont {Schuster}}, \bibinfo {author}
  {\bibfnamefont {J.}~\bibnamefont {Majer}}, \bibinfo {author} {\bibfnamefont
  {A.}~\bibnamefont {Blais}}, \bibinfo {author} {\bibfnamefont {M.~H.}\
  \bibnamefont {Devoret}}, \bibinfo {author} {\bibfnamefont {S.~M.}\
  \bibnamefont {Girvin}},\ and\ \bibinfo {author} {\bibfnamefont {R.~J.}\
  \bibnamefont {Schoelkopf}},\ }\href
  {https://doi.org/10.1103/PhysRevA.76.042319} {\bibfield  {journal} {\bibinfo
  {journal} {Phys. Rev. A}\ }\textbf {\bibinfo {volume} {76}},\ \bibinfo
  {pages} {042319} (\bibinfo {year} {2007})}\BibitemShut {NoStop}%
\bibitem [{\citenamefont {Schreier}\ \emph {et~al.}(2008)\citenamefont
  {Schreier}, \citenamefont {Houck}, \citenamefont {Koch}, \citenamefont
  {Schuster}, \citenamefont {Johnson}, \citenamefont {Chow}, \citenamefont
  {Gambetta}, \citenamefont {Majer}, \citenamefont {Frunzio}, \citenamefont
  {Devoret}, \citenamefont {Girvin},\ and\ \citenamefont
  {Schoelkopf}}]{Transmon2}%
  \BibitemOpen
  \bibfield  {author} {\bibinfo {author} {\bibfnamefont {J.~A.}\ \bibnamefont
  {Schreier}}, \bibinfo {author} {\bibfnamefont {A.~A.}\ \bibnamefont {Houck}},
  \bibinfo {author} {\bibfnamefont {J.}~\bibnamefont {Koch}}, \bibinfo {author}
  {\bibfnamefont {D.~I.}\ \bibnamefont {Schuster}}, \bibinfo {author}
  {\bibfnamefont {B.~R.}\ \bibnamefont {Johnson}}, \bibinfo {author}
  {\bibfnamefont {J.~M.}\ \bibnamefont {Chow}}, \bibinfo {author}
  {\bibfnamefont {J.~M.}\ \bibnamefont {Gambetta}}, \bibinfo {author}
  {\bibfnamefont {J.}~\bibnamefont {Majer}}, \bibinfo {author} {\bibfnamefont
  {L.}~\bibnamefont {Frunzio}}, \bibinfo {author} {\bibfnamefont {M.~H.}\
  \bibnamefont {Devoret}}, \bibinfo {author} {\bibfnamefont {S.~M.}\
  \bibnamefont {Girvin}},\ and\ \bibinfo {author} {\bibfnamefont {R.~J.}\
  \bibnamefont {Schoelkopf}},\ }\href
  {https://doi.org/10.1103/PhysRevB.77.180502} {\bibfield  {journal} {\bibinfo
  {journal} {Phys. Rev. B}\ }\textbf {\bibinfo {volume} {77}},\ \bibinfo
  {pages} {180502} (\bibinfo {year} {2008})}\BibitemShut {NoStop}%
\bibitem [{\citenamefont {Barends}\ \emph {et~al.}(2013)\citenamefont
  {Barends}, \citenamefont {Kelly}, \citenamefont {Megrant}, \citenamefont
  {Sank}, \citenamefont {Jeffrey}, \citenamefont {Chen}, \citenamefont {Yin},
  \citenamefont {Chiaro}, \citenamefont {Mutus}, \citenamefont {Neill},
  \citenamefont {O'Malley}, \citenamefont {Roushan}, \citenamefont {Wenner},
  \citenamefont {White}, \citenamefont {Cleland},\ and\ \citenamefont
  {Martinis}}]{xmon}%
  \BibitemOpen
  \bibfield  {author} {\bibinfo {author} {\bibfnamefont {R.}~\bibnamefont
  {Barends}}, \bibinfo {author} {\bibfnamefont {J.}~\bibnamefont {Kelly}},
  \bibinfo {author} {\bibfnamefont {A.}~\bibnamefont {Megrant}}, \bibinfo
  {author} {\bibfnamefont {D.}~\bibnamefont {Sank}}, \bibinfo {author}
  {\bibfnamefont {E.}~\bibnamefont {Jeffrey}}, \bibinfo {author} {\bibfnamefont
  {Y.}~\bibnamefont {Chen}}, \bibinfo {author} {\bibfnamefont {Y.}~\bibnamefont
  {Yin}}, \bibinfo {author} {\bibfnamefont {B.}~\bibnamefont {Chiaro}},
  \bibinfo {author} {\bibfnamefont {J.}~\bibnamefont {Mutus}}, \bibinfo
  {author} {\bibfnamefont {C.}~\bibnamefont {Neill}}, \bibinfo {author}
  {\bibfnamefont {P.}~\bibnamefont {O'Malley}}, \bibinfo {author}
  {\bibfnamefont {P.}~\bibnamefont {Roushan}}, \bibinfo {author} {\bibfnamefont
  {J.}~\bibnamefont {Wenner}}, \bibinfo {author} {\bibfnamefont {T.~C.}\
  \bibnamefont {White}}, \bibinfo {author} {\bibfnamefont {A.~N.}\ \bibnamefont
  {Cleland}},\ and\ \bibinfo {author} {\bibfnamefont {J.~M.}\ \bibnamefont
  {Martinis}},\ }\href {https://doi.org/10.1103/PhysRevLett.111.080502}
  {\bibfield  {journal} {\bibinfo  {journal} {Phys. Rev. Lett.}\ }\textbf
  {\bibinfo {volume} {111}},\ \bibinfo {pages} {080502} (\bibinfo {year}
  {2013})}\BibitemShut {NoStop}%
\bibitem [{\citenamefont {Vion}\ \emph {et~al.}(2002)\citenamefont {Vion},
  \citenamefont {Aassime}, \citenamefont {Cottet}, \citenamefont {Joyez},
  \citenamefont {Pothier}, \citenamefont {Urbina}, \citenamefont {Esteve},\
  and\ \citenamefont {Devoret}}]{quantronium}%
  \BibitemOpen
  \bibfield  {author} {\bibinfo {author} {\bibfnamefont {D.}~\bibnamefont
  {Vion}}, \bibinfo {author} {\bibfnamefont {A.}~\bibnamefont {Aassime}},
  \bibinfo {author} {\bibfnamefont {A.}~\bibnamefont {Cottet}}, \bibinfo
  {author} {\bibfnamefont {P.}~\bibnamefont {Joyez}}, \bibinfo {author}
  {\bibfnamefont {H.}~\bibnamefont {Pothier}}, \bibinfo {author} {\bibfnamefont
  {C.}~\bibnamefont {Urbina}}, \bibinfo {author} {\bibfnamefont
  {D.}~\bibnamefont {Esteve}},\ and\ \bibinfo {author} {\bibfnamefont {M.~H.}\
  \bibnamefont {Devoret}},\ }\href {https://doi.org/10.1126/science.1069372}
  {\bibfield  {journal} {\bibinfo  {journal} {Science}\ }\textbf {\bibinfo
  {volume} {296}},\ \bibinfo {pages} {886} (\bibinfo {year}
  {2002})}\BibitemShut {NoStop}%
\bibitem [{\citenamefont {Manucharyan}\ \emph {et~al.}(2009)\citenamefont
  {Manucharyan}, \citenamefont {Koch}, \citenamefont {Glazman},\ and\
  \citenamefont {Devoret}}]{fluxonium}%
  \BibitemOpen
  \bibfield  {author} {\bibinfo {author} {\bibfnamefont {V.~E.}\ \bibnamefont
  {Manucharyan}}, \bibinfo {author} {\bibfnamefont {J.}~\bibnamefont {Koch}},
  \bibinfo {author} {\bibfnamefont {L.~I.}\ \bibnamefont {Glazman}},\ and\
  \bibinfo {author} {\bibfnamefont {M.~H.}\ \bibnamefont {Devoret}},\ }\href
  {https://doi.org/10.1126/science.1175552} {\bibfield  {journal} {\bibinfo
  {journal} {Science}\ }\textbf {\bibinfo {volume} {326}},\ \bibinfo {pages}
  {113} (\bibinfo {year} {2009})}\BibitemShut {NoStop}%
\bibitem [{\citenamefont {Chiorescu}\ \emph {et~al.}(2003)\citenamefont
  {Chiorescu}, \citenamefont {Nakamura}, \citenamefont {Harmans},\ and\
  \citenamefont {Mooij}}]{more1}%
  \BibitemOpen
  \bibfield  {author} {\bibinfo {author} {\bibfnamefont {I.}~\bibnamefont
  {Chiorescu}}, \bibinfo {author} {\bibfnamefont {Y.}~\bibnamefont {Nakamura}},
  \bibinfo {author} {\bibfnamefont {C.~J. P.~M.}\ \bibnamefont {Harmans}},\
  and\ \bibinfo {author} {\bibfnamefont {J.~E.}\ \bibnamefont {Mooij}},\ }\href
  {https://doi.org/10.1126/science.1081045} {\bibfield  {journal} {\bibinfo
  {journal} {Science}\ }\textbf {\bibinfo {volume} {299}},\ \bibinfo {pages}
  {1869} (\bibinfo {year} {2003})}\BibitemShut {NoStop}%
\bibitem [{\citenamefont {Paik}\ \emph {et~al.}(2011)\citenamefont {Paik},
  \citenamefont {Schuster}, \citenamefont {Bishop}, \citenamefont {Kirchmair},
  \citenamefont {Catelani}, \citenamefont {Sears}, \citenamefont {Johnson},
  \citenamefont {Reagor}, \citenamefont {Frunzio}, \citenamefont {Glazman},
  \citenamefont {Girvin}, \citenamefont {Devoret},\ and\ \citenamefont
  {Schoelkopf}}]{Transmon3}%
  \BibitemOpen
  \bibfield  {author} {\bibinfo {author} {\bibfnamefont {H.}~\bibnamefont
  {Paik}}, \bibinfo {author} {\bibfnamefont {D.~I.}\ \bibnamefont {Schuster}},
  \bibinfo {author} {\bibfnamefont {L.~S.}\ \bibnamefont {Bishop}}, \bibinfo
  {author} {\bibfnamefont {G.}~\bibnamefont {Kirchmair}}, \bibinfo {author}
  {\bibfnamefont {G.}~\bibnamefont {Catelani}}, \bibinfo {author}
  {\bibfnamefont {A.~P.}\ \bibnamefont {Sears}}, \bibinfo {author}
  {\bibfnamefont {B.~R.}\ \bibnamefont {Johnson}}, \bibinfo {author}
  {\bibfnamefont {M.~J.}\ \bibnamefont {Reagor}}, \bibinfo {author}
  {\bibfnamefont {L.}~\bibnamefont {Frunzio}}, \bibinfo {author} {\bibfnamefont
  {L.~I.}\ \bibnamefont {Glazman}}, \bibinfo {author} {\bibfnamefont {S.~M.}\
  \bibnamefont {Girvin}}, \bibinfo {author} {\bibfnamefont {M.~H.}\
  \bibnamefont {Devoret}},\ and\ \bibinfo {author} {\bibfnamefont {R.~J.}\
  \bibnamefont {Schoelkopf}},\ }\href
  {https://doi.org/10.1103/PhysRevLett.107.240501} {\bibfield  {journal}
  {\bibinfo  {journal} {Phys. Rev. Lett.}\ }\textbf {\bibinfo {volume} {107}},\
  \bibinfo {pages} {240501} (\bibinfo {year} {2011})}\BibitemShut {NoStop}%
\bibitem [{\citenamefont {Hyypp\"{a}}\ \emph {et~al.}(2022)\citenamefont
  {Hyypp\"{a}}, \citenamefont {Kundu}, \citenamefont {Chan}, \citenamefont
  {Gunyh{\'{o}}}, \citenamefont {Hotari}, \citenamefont {Janzso}, \citenamefont
  {Juliusson}, \citenamefont {Kiuru}, \citenamefont {Kotilahti}, \citenamefont
  {Landra}, \citenamefont {Liu}, \citenamefont {Marxer}, \citenamefont
  {M\"{a}kinen}, \citenamefont {Orgiazzi}, \citenamefont {Palma}, \citenamefont
  {Savytskyi}, \citenamefont {Tosto}, \citenamefont {Tuorila}, \citenamefont
  {Vadimov}, \citenamefont {Li}, \citenamefont {Ockeloen-Korppi}, \citenamefont
  {Heinsoo}, \citenamefont {Tan}, \citenamefont {Hassel},\ and\ \citenamefont
  {M\"{o}tt\"{o}nen}}]{more2}%
  \BibitemOpen
  \bibfield  {author} {\bibinfo {author} {\bibfnamefont {E.}~\bibnamefont
  {Hyypp\"{a}}}, \bibinfo {author} {\bibfnamefont {S.}~\bibnamefont {Kundu}},
  \bibinfo {author} {\bibfnamefont {C.~F.}\ \bibnamefont {Chan}}, \bibinfo
  {author} {\bibfnamefont {A.}~\bibnamefont {Gunyh{\'{o}}}}, \bibinfo {author}
  {\bibfnamefont {J.}~\bibnamefont {Hotari}}, \bibinfo {author} {\bibfnamefont
  {D.}~\bibnamefont {Janzso}}, \bibinfo {author} {\bibfnamefont
  {K.}~\bibnamefont {Juliusson}}, \bibinfo {author} {\bibfnamefont
  {O.}~\bibnamefont {Kiuru}}, \bibinfo {author} {\bibfnamefont
  {J.}~\bibnamefont {Kotilahti}}, \bibinfo {author} {\bibfnamefont
  {A.}~\bibnamefont {Landra}}, \bibinfo {author} {\bibfnamefont
  {W.}~\bibnamefont {Liu}}, \bibinfo {author} {\bibfnamefont {F.}~\bibnamefont
  {Marxer}}, \bibinfo {author} {\bibfnamefont {A.}~\bibnamefont {M\"{a}kinen}},
  \bibinfo {author} {\bibfnamefont {J.-L.}\ \bibnamefont {Orgiazzi}}, \bibinfo
  {author} {\bibfnamefont {M.}~\bibnamefont {Palma}}, \bibinfo {author}
  {\bibfnamefont {M.}~\bibnamefont {Savytskyi}}, \bibinfo {author}
  {\bibfnamefont {F.}~\bibnamefont {Tosto}}, \bibinfo {author} {\bibfnamefont
  {J.}~\bibnamefont {Tuorila}}, \bibinfo {author} {\bibfnamefont
  {V.}~\bibnamefont {Vadimov}}, \bibinfo {author} {\bibfnamefont
  {T.}~\bibnamefont {Li}}, \bibinfo {author} {\bibfnamefont {C.}~\bibnamefont
  {Ockeloen-Korppi}}, \bibinfo {author} {\bibfnamefont {J.}~\bibnamefont
  {Heinsoo}}, \bibinfo {author} {\bibfnamefont {K.~Y.}\ \bibnamefont {Tan}},
  \bibinfo {author} {\bibfnamefont {J.}~\bibnamefont {Hassel}},\ and\ \bibinfo
  {author} {\bibfnamefont {M.}~\bibnamefont {M\"{o}tt\"{o}nen}},\ }\href
  {https://doi.org/10.1038/s41467-022-34614-w} {\bibfield  {journal} {\bibinfo
  {journal} {Nature Communications}\ }\textbf {\bibinfo {volume} {13}},\
  \bibinfo {pages} {6895} (\bibinfo {year} {2022})}\BibitemShut {NoStop}%
\bibitem [{\citenamefont {Martinis}\ \emph {et~al.}(2002)\citenamefont
  {Martinis}, \citenamefont {Nam}, \citenamefont {Aumentado},\ and\
  \citenamefont {Urbina}}]{more3}%
  \BibitemOpen
  \bibfield  {author} {\bibinfo {author} {\bibfnamefont {J.~M.}\ \bibnamefont
  {Martinis}}, \bibinfo {author} {\bibfnamefont {S.}~\bibnamefont {Nam}},
  \bibinfo {author} {\bibfnamefont {J.}~\bibnamefont {Aumentado}},\ and\
  \bibinfo {author} {\bibfnamefont {C.}~\bibnamefont {Urbina}},\ }\href
  {https://doi.org/10.1103/PhysRevLett.89.117901} {\bibfield  {journal}
  {\bibinfo  {journal} {Phys. Rev. Lett.}\ }\textbf {\bibinfo {volume} {89}},\
  \bibinfo {pages} {117901} (\bibinfo {year} {2002})}\BibitemShut {NoStop}%
\bibitem [{\citenamefont {Preskill}(2018)}]{nisq}%
  \BibitemOpen
  \bibfield  {author} {\bibinfo {author} {\bibfnamefont {J.}~\bibnamefont
  {Preskill}},\ }\href {https://doi.org/10.22331/q-2018-08-06-79} {\bibfield
  {journal} {\bibinfo  {journal} {{Quantum}}\ }\textbf {\bibinfo {volume}
  {2}},\ \bibinfo {pages} {79} (\bibinfo {year} {2018})}\BibitemShut {NoStop}%
\bibitem [{\citenamefont {Arute}\ \emph {et~al.}(2019)\citenamefont {Arute},
  \citenamefont {Arya}, \citenamefont {Babbush}, \citenamefont {Bacon},
  \citenamefont {Bardin}, \citenamefont {Barends}, \citenamefont {Biswas},
  \citenamefont {Boixo}, \citenamefont {Brandao}, \citenamefont {Buell},
  \citenamefont {Burkett}, \citenamefont {Chen}, \citenamefont {Chen},
  \citenamefont {Chiaro}, \citenamefont {Collins}, \citenamefont {Courtney},
  \citenamefont {Dunsworth}, \citenamefont {Farhi}, \citenamefont {Foxen},
  \citenamefont {Fowler}, \citenamefont {Gidney}, \citenamefont {Giustina},
  \citenamefont {Graff}, \citenamefont {Guerin}, \citenamefont {Habegger},
  \citenamefont {Harrigan}, \citenamefont {Hartmann}, \citenamefont {Ho},
  \citenamefont {Hoffmann}, \citenamefont {Huang}, \citenamefont {Humble},
  \citenamefont {Isakov}, \citenamefont {Jeffrey}, \citenamefont {Jiang},
  \citenamefont {Kafri}, \citenamefont {Kechedzhi}, \citenamefont {Kelly},
  \citenamefont {Klimov}, \citenamefont {Knysh}, \citenamefont {Korotkov},
  \citenamefont {Kostritsa}, \citenamefont {Landhuis}, \citenamefont
  {Lindmark}, \citenamefont {Lucero}, \citenamefont {Lyakh}, \citenamefont
  {Mandr{\`{a}}}, \citenamefont {McClean}, \citenamefont {McEwen},
  \citenamefont {Megrant}, \citenamefont {Mi}, \citenamefont {Michielsen},
  \citenamefont {Mohseni}, \citenamefont {Mutus}, \citenamefont {Naaman},
  \citenamefont {Neeley}, \citenamefont {Neill}, \citenamefont {Niu},
  \citenamefont {Ostby}, \citenamefont {Petukhov}, \citenamefont {Platt},
  \citenamefont {Quintana}, \citenamefont {Rieffel}, \citenamefont {Roushan},
  \citenamefont {Rubin}, \citenamefont {Sank}, \citenamefont {Satzinger},
  \citenamefont {Smelyanskiy}, \citenamefont {Sung}, \citenamefont
  {Trevithick}, \citenamefont {Vainsencher}, \citenamefont {Villalonga},
  \citenamefont {White}, \citenamefont {Yao}, \citenamefont {Yeh},
  \citenamefont {Zalcman}, \citenamefont {Neven},\ and\ \citenamefont
  {Martinis}}]{qsup1}%
  \BibitemOpen
  \bibfield  {author} {\bibinfo {author} {\bibfnamefont {F.}~\bibnamefont
  {Arute}}, \bibinfo {author} {\bibfnamefont {K.}~\bibnamefont {Arya}},
  \bibinfo {author} {\bibfnamefont {R.}~\bibnamefont {Babbush}}, \bibinfo
  {author} {\bibfnamefont {D.}~\bibnamefont {Bacon}}, \bibinfo {author}
  {\bibfnamefont {J.~C.}\ \bibnamefont {Bardin}}, \bibinfo {author}
  {\bibfnamefont {R.}~\bibnamefont {Barends}}, \bibinfo {author} {\bibfnamefont
  {R.}~\bibnamefont {Biswas}}, \bibinfo {author} {\bibfnamefont
  {S.}~\bibnamefont {Boixo}}, \bibinfo {author} {\bibfnamefont {F.~G. S.~L.}\
  \bibnamefont {Brandao}}, \bibinfo {author} {\bibfnamefont {D.~A.}\
  \bibnamefont {Buell}}, \bibinfo {author} {\bibfnamefont {B.}~\bibnamefont
  {Burkett}}, \bibinfo {author} {\bibfnamefont {Y.}~\bibnamefont {Chen}},
  \bibinfo {author} {\bibfnamefont {Z.}~\bibnamefont {Chen}}, \bibinfo {author}
  {\bibfnamefont {B.}~\bibnamefont {Chiaro}}, \bibinfo {author} {\bibfnamefont
  {R.}~\bibnamefont {Collins}}, \bibinfo {author} {\bibfnamefont
  {W.}~\bibnamefont {Courtney}}, \bibinfo {author} {\bibfnamefont
  {A.}~\bibnamefont {Dunsworth}}, \bibinfo {author} {\bibfnamefont
  {E.}~\bibnamefont {Farhi}}, \bibinfo {author} {\bibfnamefont
  {B.}~\bibnamefont {Foxen}}, \bibinfo {author} {\bibfnamefont
  {A.}~\bibnamefont {Fowler}}, \bibinfo {author} {\bibfnamefont
  {C.}~\bibnamefont {Gidney}}, \bibinfo {author} {\bibfnamefont
  {M.}~\bibnamefont {Giustina}}, \bibinfo {author} {\bibfnamefont
  {R.}~\bibnamefont {Graff}}, \bibinfo {author} {\bibfnamefont
  {K.}~\bibnamefont {Guerin}}, \bibinfo {author} {\bibfnamefont
  {S.}~\bibnamefont {Habegger}}, \bibinfo {author} {\bibfnamefont {M.~P.}\
  \bibnamefont {Harrigan}}, \bibinfo {author} {\bibfnamefont {M.~J.}\
  \bibnamefont {Hartmann}}, \bibinfo {author} {\bibfnamefont {A.}~\bibnamefont
  {Ho}}, \bibinfo {author} {\bibfnamefont {M.}~\bibnamefont {Hoffmann}},
  \bibinfo {author} {\bibfnamefont {T.}~\bibnamefont {Huang}}, \bibinfo
  {author} {\bibfnamefont {T.~S.}\ \bibnamefont {Humble}}, \bibinfo {author}
  {\bibfnamefont {S.~V.}\ \bibnamefont {Isakov}}, \bibinfo {author}
  {\bibfnamefont {E.}~\bibnamefont {Jeffrey}}, \bibinfo {author} {\bibfnamefont
  {Z.}~\bibnamefont {Jiang}}, \bibinfo {author} {\bibfnamefont
  {D.}~\bibnamefont {Kafri}}, \bibinfo {author} {\bibfnamefont
  {K.}~\bibnamefont {Kechedzhi}}, \bibinfo {author} {\bibfnamefont
  {J.}~\bibnamefont {Kelly}}, \bibinfo {author} {\bibfnamefont {P.~V.}\
  \bibnamefont {Klimov}}, \bibinfo {author} {\bibfnamefont {S.}~\bibnamefont
  {Knysh}}, \bibinfo {author} {\bibfnamefont {A.}~\bibnamefont {Korotkov}},
  \bibinfo {author} {\bibfnamefont {F.}~\bibnamefont {Kostritsa}}, \bibinfo
  {author} {\bibfnamefont {D.}~\bibnamefont {Landhuis}}, \bibinfo {author}
  {\bibfnamefont {M.}~\bibnamefont {Lindmark}}, \bibinfo {author}
  {\bibfnamefont {E.}~\bibnamefont {Lucero}}, \bibinfo {author} {\bibfnamefont
  {D.}~\bibnamefont {Lyakh}}, \bibinfo {author} {\bibfnamefont
  {S.}~\bibnamefont {Mandr{\`{a}}}}, \bibinfo {author} {\bibfnamefont {J.~R.}\
  \bibnamefont {McClean}}, \bibinfo {author} {\bibfnamefont {M.}~\bibnamefont
  {McEwen}}, \bibinfo {author} {\bibfnamefont {A.}~\bibnamefont {Megrant}},
  \bibinfo {author} {\bibfnamefont {X.}~\bibnamefont {Mi}}, \bibinfo {author}
  {\bibfnamefont {K.}~\bibnamefont {Michielsen}}, \bibinfo {author}
  {\bibfnamefont {M.}~\bibnamefont {Mohseni}}, \bibinfo {author} {\bibfnamefont
  {J.}~\bibnamefont {Mutus}}, \bibinfo {author} {\bibfnamefont
  {O.}~\bibnamefont {Naaman}}, \bibinfo {author} {\bibfnamefont
  {M.}~\bibnamefont {Neeley}}, \bibinfo {author} {\bibfnamefont
  {C.}~\bibnamefont {Neill}}, \bibinfo {author} {\bibfnamefont {M.~Y.}\
  \bibnamefont {Niu}}, \bibinfo {author} {\bibfnamefont {E.}~\bibnamefont
  {Ostby}}, \bibinfo {author} {\bibfnamefont {A.}~\bibnamefont {Petukhov}},
  \bibinfo {author} {\bibfnamefont {J.~C.}\ \bibnamefont {Platt}}, \bibinfo
  {author} {\bibfnamefont {C.}~\bibnamefont {Quintana}}, \bibinfo {author}
  {\bibfnamefont {E.~G.}\ \bibnamefont {Rieffel}}, \bibinfo {author}
  {\bibfnamefont {P.}~\bibnamefont {Roushan}}, \bibinfo {author} {\bibfnamefont
  {N.~C.}\ \bibnamefont {Rubin}}, \bibinfo {author} {\bibfnamefont
  {D.}~\bibnamefont {Sank}}, \bibinfo {author} {\bibfnamefont {K.~J.}\
  \bibnamefont {Satzinger}}, \bibinfo {author} {\bibfnamefont {V.}~\bibnamefont
  {Smelyanskiy}}, \bibinfo {author} {\bibfnamefont {K.~J.}\ \bibnamefont
  {Sung}}, \bibinfo {author} {\bibfnamefont {M.~D.}\ \bibnamefont
  {Trevithick}}, \bibinfo {author} {\bibfnamefont {A.}~\bibnamefont
  {Vainsencher}}, \bibinfo {author} {\bibfnamefont {B.}~\bibnamefont
  {Villalonga}}, \bibinfo {author} {\bibfnamefont {T.}~\bibnamefont {White}},
  \bibinfo {author} {\bibfnamefont {Z.~J.}\ \bibnamefont {Yao}}, \bibinfo
  {author} {\bibfnamefont {P.}~\bibnamefont {Yeh}}, \bibinfo {author}
  {\bibfnamefont {A.}~\bibnamefont {Zalcman}}, \bibinfo {author} {\bibfnamefont
  {H.}~\bibnamefont {Neven}},\ and\ \bibinfo {author} {\bibfnamefont {J.~M.}\
  \bibnamefont {Martinis}},\ }\href {https://doi.org/10.1038/s41586-019-1666-5}
  {\bibfield  {journal} {\bibinfo  {journal} {Nature}\ }\textbf {\bibinfo
  {volume} {574}},\ \bibinfo {pages} {505} (\bibinfo {year}
  {2019})}\BibitemShut {NoStop}%
\bibitem [{\citenamefont {Bouchiat}\ \emph {et~al.}(1998)\citenamefont
  {Bouchiat}, \citenamefont {Vion}, \citenamefont {Joyez}, \citenamefont
  {Esteve},\ and\ \citenamefont {Devoret}}]{Bouchiat1998}%
  \BibitemOpen
  \bibfield  {author} {\bibinfo {author} {\bibfnamefont {V.}~\bibnamefont
  {Bouchiat}}, \bibinfo {author} {\bibfnamefont {D.}~\bibnamefont {Vion}},
  \bibinfo {author} {\bibfnamefont {P.}~\bibnamefont {Joyez}}, \bibinfo
  {author} {\bibfnamefont {D.}~\bibnamefont {Esteve}},\ and\ \bibinfo {author}
  {\bibfnamefont {M.~H.}\ \bibnamefont {Devoret}},\ }\href
  {https://doi.org/10.1238/Physica.Topical.076a00165} {\bibfield  {journal}
  {\bibinfo  {journal} {Physica Scripta}\ }\textbf {\bibinfo {volume} {1998}},\
  \bibinfo {pages} {165} (\bibinfo {year} {1998})}\BibitemShut {NoStop}%
\bibitem [{\citenamefont {Nakamura}\ \emph {et~al.}(1999)\citenamefont
  {Nakamura}, \citenamefont {Pashkin},\ and\ \citenamefont
  {Tsai}}]{Nakamura1999}%
  \BibitemOpen
  \bibfield  {author} {\bibinfo {author} {\bibfnamefont {Y.}~\bibnamefont
  {Nakamura}}, \bibinfo {author} {\bibfnamefont {Y.~A.}\ \bibnamefont
  {Pashkin}},\ and\ \bibinfo {author} {\bibfnamefont {J.~S.}\ \bibnamefont
  {Tsai}},\ }\href {https://doi.org/10.1038/19718} {\bibfield  {journal}
  {\bibinfo  {journal} {Nature}\ }\textbf {\bibinfo {volume} {398}},\ \bibinfo
  {pages} {786} (\bibinfo {year} {1999})}\BibitemShut {NoStop}%
\bibitem [{\citenamefont {Wang}\ \emph {et~al.}(2015)\citenamefont {Wang},
  \citenamefont {Axline}, \citenamefont {Gao}, \citenamefont {Brecht},
  \citenamefont {Chu}, \citenamefont {Frunzio}, \citenamefont {Devoret},\ and\
  \citenamefont {Schoelkopf}}]{Dielectric}%
  \BibitemOpen
  \bibfield  {author} {\bibinfo {author} {\bibfnamefont {C.}~\bibnamefont
  {Wang}}, \bibinfo {author} {\bibfnamefont {C.}~\bibnamefont {Axline}},
  \bibinfo {author} {\bibfnamefont {Y.~Y.}\ \bibnamefont {Gao}}, \bibinfo
  {author} {\bibfnamefont {T.}~\bibnamefont {Brecht}}, \bibinfo {author}
  {\bibfnamefont {Y.}~\bibnamefont {Chu}}, \bibinfo {author} {\bibfnamefont
  {L.}~\bibnamefont {Frunzio}}, \bibinfo {author} {\bibfnamefont {M.~H.}\
  \bibnamefont {Devoret}},\ and\ \bibinfo {author} {\bibfnamefont {R.~J.}\
  \bibnamefont {Schoelkopf}},\ }\href {https://doi.org/10.1063/1.4934486}
  {\bibfield  {journal} {\bibinfo  {journal} {Applied Physics Letters}\
  }\textbf {\bibinfo {volume} {107}},\ \bibinfo {pages} {162601} (\bibinfo
  {year} {2015})}\BibitemShut {NoStop}%
\bibitem [{\citenamefont {Dial}\ \emph {et~al.}(2016)\citenamefont {Dial},
  \citenamefont {McClure}, \citenamefont {Poletto}, \citenamefont {Keefe},
  \citenamefont {Rothwell}, \citenamefont {Gambetta}, \citenamefont {Abraham},
  \citenamefont {Chow},\ and\ \citenamefont {Steffen}}]{Dial2016}%
  \BibitemOpen
  \bibfield  {author} {\bibinfo {author} {\bibfnamefont {O.}~\bibnamefont
  {Dial}}, \bibinfo {author} {\bibfnamefont {D.~T.}\ \bibnamefont {McClure}},
  \bibinfo {author} {\bibfnamefont {S.}~\bibnamefont {Poletto}}, \bibinfo
  {author} {\bibfnamefont {G.~A.}\ \bibnamefont {Keefe}}, \bibinfo {author}
  {\bibfnamefont {M.~B.}\ \bibnamefont {Rothwell}}, \bibinfo {author}
  {\bibfnamefont {J.~M.}\ \bibnamefont {Gambetta}}, \bibinfo {author}
  {\bibfnamefont {D.~W.}\ \bibnamefont {Abraham}}, \bibinfo {author}
  {\bibfnamefont {J.~M.}\ \bibnamefont {Chow}},\ and\ \bibinfo {author}
  {\bibfnamefont {M.}~\bibnamefont {Steffen}},\ }\href
  {https://doi.org/10.1088/0953-2048/29/4/044001} {\bibfield  {journal}
  {\bibinfo  {journal} {Superconductor Science and Technology}\ }\textbf
  {\bibinfo {volume} {29}},\ \bibinfo {pages} {044001} (\bibinfo {year}
  {2016})}\BibitemShut {NoStop}%
\bibitem [{\citenamefont {Veps\"{a}l\"{a}inen}\ \emph
  {et~al.}(2020)\citenamefont {Veps\"{a}l\"{a}inen}, \citenamefont {Karamlou},
  \citenamefont {Orrell}, \citenamefont {Dogra}, \citenamefont {Loer},
  \citenamefont {Vasconcelos}, \citenamefont {Kim}, \citenamefont {Melville},
  \citenamefont {Niedzielski}, \citenamefont {Yoder}, \citenamefont
  {Gustavsson}, \citenamefont {Formaggio}, \citenamefont {VanDevender},\ and\
  \citenamefont {Oliver}}]{Vepslinen2020}%
  \BibitemOpen
  \bibfield  {author} {\bibinfo {author} {\bibfnamefont {A.~P.}\ \bibnamefont
  {Veps\"{a}l\"{a}inen}}, \bibinfo {author} {\bibfnamefont {A.~H.}\
  \bibnamefont {Karamlou}}, \bibinfo {author} {\bibfnamefont {J.~L.}\
  \bibnamefont {Orrell}}, \bibinfo {author} {\bibfnamefont {A.~S.}\
  \bibnamefont {Dogra}}, \bibinfo {author} {\bibfnamefont {B.}~\bibnamefont
  {Loer}}, \bibinfo {author} {\bibfnamefont {F.}~\bibnamefont {Vasconcelos}},
  \bibinfo {author} {\bibfnamefont {D.~K.}\ \bibnamefont {Kim}}, \bibinfo
  {author} {\bibfnamefont {A.~J.}\ \bibnamefont {Melville}}, \bibinfo {author}
  {\bibfnamefont {B.~M.}\ \bibnamefont {Niedzielski}}, \bibinfo {author}
  {\bibfnamefont {J.~L.}\ \bibnamefont {Yoder}}, \bibinfo {author}
  {\bibfnamefont {S.}~\bibnamefont {Gustavsson}}, \bibinfo {author}
  {\bibfnamefont {J.~A.}\ \bibnamefont {Formaggio}}, \bibinfo {author}
  {\bibfnamefont {B.~A.}\ \bibnamefont {VanDevender}},\ and\ \bibinfo {author}
  {\bibfnamefont {W.~D.}\ \bibnamefont {Oliver}},\ }\href
  {https://doi.org/10.1038/s41586-020-2619-8} {\bibfield  {journal} {\bibinfo
  {journal} {Nature}\ }\textbf {\bibinfo {volume} {584}},\ \bibinfo {pages}
  {551} (\bibinfo {year} {2020})}\BibitemShut {NoStop}%
\bibitem [{\citenamefont {Wilen}\ \emph {et~al.}(2021)\citenamefont {Wilen},
  \citenamefont {Abdullah}, \citenamefont {Kurinsky}, \citenamefont {Stanford},
  \citenamefont {Cardani}, \citenamefont {D'Imperio}, \citenamefont {Tomei},
  \citenamefont {Faoro}, \citenamefont {Ioffe}, \citenamefont {Liu},
  \citenamefont {Opremcak}, \citenamefont {Christensen}, \citenamefont
  {DuBois},\ and\ \citenamefont {McDermott}}]{Wilen2021}%
  \BibitemOpen
  \bibfield  {author} {\bibinfo {author} {\bibfnamefont {C.~D.}\ \bibnamefont
  {Wilen}}, \bibinfo {author} {\bibfnamefont {S.}~\bibnamefont {Abdullah}},
  \bibinfo {author} {\bibfnamefont {N.~A.}\ \bibnamefont {Kurinsky}}, \bibinfo
  {author} {\bibfnamefont {C.}~\bibnamefont {Stanford}}, \bibinfo {author}
  {\bibfnamefont {L.}~\bibnamefont {Cardani}}, \bibinfo {author} {\bibfnamefont
  {G.}~\bibnamefont {D'Imperio}}, \bibinfo {author} {\bibfnamefont
  {C.}~\bibnamefont {Tomei}}, \bibinfo {author} {\bibfnamefont
  {L.}~\bibnamefont {Faoro}}, \bibinfo {author} {\bibfnamefont {L.~B.}\
  \bibnamefont {Ioffe}}, \bibinfo {author} {\bibfnamefont {C.~H.}\ \bibnamefont
  {Liu}}, \bibinfo {author} {\bibfnamefont {A.}~\bibnamefont {Opremcak}},
  \bibinfo {author} {\bibfnamefont {B.~G.}\ \bibnamefont {Christensen}},
  \bibinfo {author} {\bibfnamefont {J.~L.}\ \bibnamefont {DuBois}},\ and\
  \bibinfo {author} {\bibfnamefont {R.}~\bibnamefont {McDermott}},\ }\href
  {https://doi.org/10.1038/s41586-021-03557-5} {\bibfield  {journal} {\bibinfo
  {journal} {Nature}\ }\textbf {\bibinfo {volume} {594}},\ \bibinfo {pages}
  {369} (\bibinfo {year} {2021})}\BibitemShut {NoStop}%
\bibitem [{\citenamefont {McEwen}\ \emph {et~al.}(2021)\citenamefont {McEwen},
  \citenamefont {Faoro}, \citenamefont {Arya}, \citenamefont {Dunsworth},
  \citenamefont {Huang}, \citenamefont {Kim}, \citenamefont {Burkett},
  \citenamefont {Fowler}, \citenamefont {Arute}, \citenamefont {Bardin},
  \citenamefont {Bengtsson}, \citenamefont {Bilmes}, \citenamefont {Buckley},
  \citenamefont {Bushnell}, \citenamefont {Chen}, \citenamefont {Collins},
  \citenamefont {Demura}, \citenamefont {Derk}, \citenamefont {Erickson},
  \citenamefont {Giustina}, \citenamefont {Harrington}, \citenamefont {Hong},
  \citenamefont {Jeffrey}, \citenamefont {Kelly}, \citenamefont {Klimov},
  \citenamefont {Kostritsa}, \citenamefont {Laptev}, \citenamefont {Locharla},
  \citenamefont {Mi}, \citenamefont {Miao}, \citenamefont {Montazeri},
  \citenamefont {Mutus}, \citenamefont {Naaman}, \citenamefont {Neeley},
  \citenamefont {Neill}, \citenamefont {Opremcak}, \citenamefont {Quintana},
  \citenamefont {Redd}, \citenamefont {Roushan}, \citenamefont {Sank},
  \citenamefont {Satzinger}, \citenamefont {Shvarts}, \citenamefont {White},
  \citenamefont {Yao}, \citenamefont {Yeh}, \citenamefont {Yoo}, \citenamefont
  {Chen}, \citenamefont {Smelyanskiy}, \citenamefont {Martinis}, \citenamefont
  {Neven}, \citenamefont {Megrant}, \citenamefont {Ioffe},\ and\ \citenamefont
  {Barends}}]{McEwen2021}%
  \BibitemOpen
  \bibfield  {author} {\bibinfo {author} {\bibfnamefont {M.}~\bibnamefont
  {McEwen}}, \bibinfo {author} {\bibfnamefont {L.}~\bibnamefont {Faoro}},
  \bibinfo {author} {\bibfnamefont {K.}~\bibnamefont {Arya}}, \bibinfo {author}
  {\bibfnamefont {A.}~\bibnamefont {Dunsworth}}, \bibinfo {author}
  {\bibfnamefont {T.}~\bibnamefont {Huang}}, \bibinfo {author} {\bibfnamefont
  {S.}~\bibnamefont {Kim}}, \bibinfo {author} {\bibfnamefont {B.}~\bibnamefont
  {Burkett}}, \bibinfo {author} {\bibfnamefont {A.}~\bibnamefont {Fowler}},
  \bibinfo {author} {\bibfnamefont {F.}~\bibnamefont {Arute}}, \bibinfo
  {author} {\bibfnamefont {J.~C.}\ \bibnamefont {Bardin}}, \bibinfo {author}
  {\bibfnamefont {A.}~\bibnamefont {Bengtsson}}, \bibinfo {author}
  {\bibfnamefont {A.}~\bibnamefont {Bilmes}}, \bibinfo {author} {\bibfnamefont
  {B.~B.}\ \bibnamefont {Buckley}}, \bibinfo {author} {\bibfnamefont
  {N.}~\bibnamefont {Bushnell}}, \bibinfo {author} {\bibfnamefont
  {Z.}~\bibnamefont {Chen}}, \bibinfo {author} {\bibfnamefont {R.}~\bibnamefont
  {Collins}}, \bibinfo {author} {\bibfnamefont {S.}~\bibnamefont {Demura}},
  \bibinfo {author} {\bibfnamefont {A.~R.}\ \bibnamefont {Derk}}, \bibinfo
  {author} {\bibfnamefont {C.}~\bibnamefont {Erickson}}, \bibinfo {author}
  {\bibfnamefont {M.}~\bibnamefont {Giustina}}, \bibinfo {author}
  {\bibfnamefont {S.~D.}\ \bibnamefont {Harrington}}, \bibinfo {author}
  {\bibfnamefont {S.}~\bibnamefont {Hong}}, \bibinfo {author} {\bibfnamefont
  {E.}~\bibnamefont {Jeffrey}}, \bibinfo {author} {\bibfnamefont
  {J.}~\bibnamefont {Kelly}}, \bibinfo {author} {\bibfnamefont {P.~V.}\
  \bibnamefont {Klimov}}, \bibinfo {author} {\bibfnamefont {F.}~\bibnamefont
  {Kostritsa}}, \bibinfo {author} {\bibfnamefont {P.}~\bibnamefont {Laptev}},
  \bibinfo {author} {\bibfnamefont {A.}~\bibnamefont {Locharla}}, \bibinfo
  {author} {\bibfnamefont {X.}~\bibnamefont {Mi}}, \bibinfo {author}
  {\bibfnamefont {K.~C.}\ \bibnamefont {Miao}}, \bibinfo {author}
  {\bibfnamefont {S.}~\bibnamefont {Montazeri}}, \bibinfo {author}
  {\bibfnamefont {J.}~\bibnamefont {Mutus}}, \bibinfo {author} {\bibfnamefont
  {O.}~\bibnamefont {Naaman}}, \bibinfo {author} {\bibfnamefont
  {M.}~\bibnamefont {Neeley}}, \bibinfo {author} {\bibfnamefont
  {C.}~\bibnamefont {Neill}}, \bibinfo {author} {\bibfnamefont
  {A.}~\bibnamefont {Opremcak}}, \bibinfo {author} {\bibfnamefont
  {C.}~\bibnamefont {Quintana}}, \bibinfo {author} {\bibfnamefont
  {N.}~\bibnamefont {Redd}}, \bibinfo {author} {\bibfnamefont {P.}~\bibnamefont
  {Roushan}}, \bibinfo {author} {\bibfnamefont {D.}~\bibnamefont {Sank}},
  \bibinfo {author} {\bibfnamefont {K.~J.}\ \bibnamefont {Satzinger}}, \bibinfo
  {author} {\bibfnamefont {V.}~\bibnamefont {Shvarts}}, \bibinfo {author}
  {\bibfnamefont {T.}~\bibnamefont {White}}, \bibinfo {author} {\bibfnamefont
  {Z.~J.}\ \bibnamefont {Yao}}, \bibinfo {author} {\bibfnamefont
  {P.}~\bibnamefont {Yeh}}, \bibinfo {author} {\bibfnamefont {J.}~\bibnamefont
  {Yoo}}, \bibinfo {author} {\bibfnamefont {Y.}~\bibnamefont {Chen}}, \bibinfo
  {author} {\bibfnamefont {V.}~\bibnamefont {Smelyanskiy}}, \bibinfo {author}
  {\bibfnamefont {J.~M.}\ \bibnamefont {Martinis}}, \bibinfo {author}
  {\bibfnamefont {H.}~\bibnamefont {Neven}}, \bibinfo {author} {\bibfnamefont
  {A.}~\bibnamefont {Megrant}}, \bibinfo {author} {\bibfnamefont
  {L.}~\bibnamefont {Ioffe}},\ and\ \bibinfo {author} {\bibfnamefont
  {R.}~\bibnamefont {Barends}},\ }\href
  {https://doi.org/10.1038/s41567-021-01432-8} {\bibfield  {journal} {\bibinfo
  {journal} {Nature Physics}\ }\textbf {\bibinfo {volume} {18}},\ \bibinfo
  {pages} {107} (\bibinfo {year} {2021})}\BibitemShut {NoStop}%
\bibitem [{\citenamefont {Houck}\ \emph {et~al.}(2008)\citenamefont {Houck},
  \citenamefont {Schreier}, \citenamefont {Johnson}, \citenamefont {Chow},
  \citenamefont {Koch}, \citenamefont {Gambetta}, \citenamefont {Schuster},
  \citenamefont {Frunzio}, \citenamefont {Devoret}, \citenamefont {Girvin},\
  and\ \citenamefont {Schoelkopf}}]{Transmon4}%
  \BibitemOpen
  \bibfield  {author} {\bibinfo {author} {\bibfnamefont {A.~A.}\ \bibnamefont
  {Houck}}, \bibinfo {author} {\bibfnamefont {J.~A.}\ \bibnamefont {Schreier}},
  \bibinfo {author} {\bibfnamefont {B.~R.}\ \bibnamefont {Johnson}}, \bibinfo
  {author} {\bibfnamefont {J.~M.}\ \bibnamefont {Chow}}, \bibinfo {author}
  {\bibfnamefont {J.}~\bibnamefont {Koch}}, \bibinfo {author} {\bibfnamefont
  {J.~M.}\ \bibnamefont {Gambetta}}, \bibinfo {author} {\bibfnamefont {D.~I.}\
  \bibnamefont {Schuster}}, \bibinfo {author} {\bibfnamefont {L.}~\bibnamefont
  {Frunzio}}, \bibinfo {author} {\bibfnamefont {M.~H.}\ \bibnamefont
  {Devoret}}, \bibinfo {author} {\bibfnamefont {S.~M.}\ \bibnamefont
  {Girvin}},\ and\ \bibinfo {author} {\bibfnamefont {R.~J.}\ \bibnamefont
  {Schoelkopf}},\ }\href {https://doi.org/10.1103/PhysRevLett.101.080502}
  {\bibfield  {journal} {\bibinfo  {journal} {Phys. Rev. Lett.}\ }\textbf
  {\bibinfo {volume} {101}},\ \bibinfo {pages} {080502} (\bibinfo {year}
  {2008})}\BibitemShut {NoStop}%
\bibitem [{\citenamefont {Unruh}(1995)}]{Unruh1995}%
  \BibitemOpen
  \bibfield  {author} {\bibinfo {author} {\bibfnamefont {W.~G.}\ \bibnamefont
  {Unruh}},\ }\href {https://doi.org/10.1103/PhysRevA.51.992} {\bibfield
  {journal} {\bibinfo  {journal} {Phys. Rev. A}\ }\textbf {\bibinfo {volume}
  {51}},\ \bibinfo {pages} {992} (\bibinfo {year} {1995})}\BibitemShut
  {NoStop}%
\bibitem [{\citenamefont {Shor}(1995)}]{Shor1995}%
  \BibitemOpen
  \bibfield  {author} {\bibinfo {author} {\bibfnamefont {P.~W.}\ \bibnamefont
  {Shor}},\ }\href {https://doi.org/10.1103/PhysRevA.52.R2493} {\bibfield
  {journal} {\bibinfo  {journal} {Phys. Rev. A}\ }\textbf {\bibinfo {volume}
  {52}},\ \bibinfo {pages} {R2493} (\bibinfo {year} {1995})}\BibitemShut
  {NoStop}%
\bibitem [{\citenamefont {Steane}(1996)}]{Steane1996}%
  \BibitemOpen
  \bibfield  {author} {\bibinfo {author} {\bibfnamefont {A.~M.}\ \bibnamefont
  {Steane}},\ }\href {https://doi.org/10.1103/PhysRevLett.77.793} {\bibfield
  {journal} {\bibinfo  {journal} {Phys. Rev. Lett.}\ }\textbf {\bibinfo
  {volume} {77}},\ \bibinfo {pages} {793} (\bibinfo {year} {1996})}\BibitemShut
  {NoStop}%
\bibitem [{\citenamefont {Devitt}\ \emph {et~al.}(2013)\citenamefont {Devitt},
  \citenamefont {Munro},\ and\ \citenamefont {Nemoto}}]{Devitt2013}%
  \BibitemOpen
  \bibfield  {author} {\bibinfo {author} {\bibfnamefont {S.~J.}\ \bibnamefont
  {Devitt}}, \bibinfo {author} {\bibfnamefont {W.~J.}\ \bibnamefont {Munro}},\
  and\ \bibinfo {author} {\bibfnamefont {K.}~\bibnamefont {Nemoto}},\ }\href
  {https://doi.org/10.1088/0034-4885/76/7/076001} {\bibfield  {journal}
  {\bibinfo  {journal} {Reports on Progress in Physics}\ }\textbf {\bibinfo
  {volume} {76}},\ \bibinfo {pages} {076001} (\bibinfo {year}
  {2013})}\BibitemShut {NoStop}%
\bibitem [{\citenamefont {Clemens}\ \emph {et~al.}(2004)\citenamefont
  {Clemens}, \citenamefont {Siddiqui},\ and\ \citenamefont
  {Gea-Banacloche}}]{QEC1}%
  \BibitemOpen
  \bibfield  {author} {\bibinfo {author} {\bibfnamefont {J.~P.}\ \bibnamefont
  {Clemens}}, \bibinfo {author} {\bibfnamefont {S.}~\bibnamefont {Siddiqui}},\
  and\ \bibinfo {author} {\bibfnamefont {J.}~\bibnamefont {Gea-Banacloche}},\
  }\href {https://doi.org/10.1103/PhysRevA.69.062313} {\bibfield  {journal}
  {\bibinfo  {journal} {Phys. Rev. A}\ }\textbf {\bibinfo {volume} {69}},\
  \bibinfo {pages} {062313} (\bibinfo {year} {2004})}\BibitemShut {NoStop}%
\bibitem [{\citenamefont {Klesse}\ and\ \citenamefont {Frank}(2005)}]{QEC2}%
  \BibitemOpen
  \bibfield  {author} {\bibinfo {author} {\bibfnamefont {R.}~\bibnamefont
  {Klesse}}\ and\ \bibinfo {author} {\bibfnamefont {S.}~\bibnamefont {Frank}},\
  }\href {https://doi.org/10.1103/PhysRevLett.95.230503} {\bibfield  {journal}
  {\bibinfo  {journal} {Phys. Rev. Lett.}\ }\textbf {\bibinfo {volume} {95}},\
  \bibinfo {pages} {230503} (\bibinfo {year} {2005})}\BibitemShut {NoStop}%
\bibitem [{\citenamefont {Aharonov}\ and\ \citenamefont
  {Ben-Or}(2008)}]{QECcorr1}%
  \BibitemOpen
  \bibfield  {author} {\bibinfo {author} {\bibfnamefont {D.}~\bibnamefont
  {Aharonov}}\ and\ \bibinfo {author} {\bibfnamefont {M.}~\bibnamefont
  {Ben-Or}},\ }\href {https://doi.org/10.1137/S0097539799359385} {\bibfield
  {journal} {\bibinfo  {journal} {SIAM Journal on Computing}\ }\textbf
  {\bibinfo {volume} {38}},\ \bibinfo {pages} {1207} (\bibinfo {year}
  {2008})}\BibitemShut {NoStop}%
\bibitem [{\citenamefont {Terhal}\ and\ \citenamefont
  {Burkard}(2005)}]{Terhal2005}%
  \BibitemOpen
  \bibfield  {author} {\bibinfo {author} {\bibfnamefont {B.~M.}\ \bibnamefont
  {Terhal}}\ and\ \bibinfo {author} {\bibfnamefont {G.}~\bibnamefont
  {Burkard}},\ }\href {https://doi.org/10.1103/PhysRevA.71.012336} {\bibfield
  {journal} {\bibinfo  {journal} {Phys. Rev. A}\ }\textbf {\bibinfo {volume}
  {71}},\ \bibinfo {pages} {012336} (\bibinfo {year} {2005})}\BibitemShut
  {NoStop}%
\bibitem [{\citenamefont {Aliferis}\ \emph {et~al.}(2006)\citenamefont
  {Aliferis}, \citenamefont {Gottesman},\ and\ \citenamefont
  {Preskill}}]{Aliferis2006}%
  \BibitemOpen
  \bibfield  {author} {\bibinfo {author} {\bibfnamefont {P.}~\bibnamefont
  {Aliferis}}, \bibinfo {author} {\bibfnamefont {D.}~\bibnamefont
  {Gottesman}},\ and\ \bibinfo {author} {\bibfnamefont {J.}~\bibnamefont
  {Preskill}},\ }\href {https://doi.org/10.26421/qic6.2-1} {\bibfield
  {journal} {\bibinfo  {journal} {Quantum Information and Computation}\
  }\textbf {\bibinfo {volume} {6}},\ \bibinfo {pages} {97} (\bibinfo {year}
  {2006})}\BibitemShut {NoStop}%
\bibitem [{\citenamefont {Preskill}(2013)}]{Preskill2013}%
  \BibitemOpen
  \bibfield  {author} {\bibinfo {author} {\bibfnamefont {J.}~\bibnamefont
  {Preskill}},\ }\href {https://doi.org/10.26421/qic13.3-4-1} {\bibfield
  {journal} {\bibinfo  {journal} {Quantum Information and Computation}\
  }\textbf {\bibinfo {volume} {13}},\ \bibinfo {pages} {181} (\bibinfo {year}
  {2013})}\BibitemShut {NoStop}%
\bibitem [{\citenamefont {Breuer}\ and\ \citenamefont
  {Petruccione}(2002)}]{BRE02}%
  \BibitemOpen
  \bibfield  {author} {\bibinfo {author} {\bibfnamefont {H.~P.}\ \bibnamefont
  {Breuer}}\ and\ \bibinfo {author} {\bibfnamefont {F.}~\bibnamefont
  {Petruccione}},\ }\href@noop {} {\emph {\bibinfo {title} {The theory of open
  quantum systems}}}\ (\bibinfo  {publisher} {Oxford University Press},\
  \bibinfo {address} {Great Clarendon Street},\ \bibinfo {year}
  {2002})\BibitemShut {NoStop}%
\bibitem [{\citenamefont {Hall}\ \emph {et~al.}(2014)\citenamefont {Hall},
  \citenamefont {Cresser}, \citenamefont {Li},\ and\ \citenamefont
  {Andersson}}]{Hall2014}%
  \BibitemOpen
  \bibfield  {author} {\bibinfo {author} {\bibfnamefont {M.~J.~W.}\
  \bibnamefont {Hall}}, \bibinfo {author} {\bibfnamefont {J.~D.}\ \bibnamefont
  {Cresser}}, \bibinfo {author} {\bibfnamefont {L.}~\bibnamefont {Li}},\ and\
  \bibinfo {author} {\bibfnamefont {E.}~\bibnamefont {Andersson}},\ }\href
  {https://doi.org/10.1103/PhysRevA.89.042120} {\bibfield  {journal} {\bibinfo
  {journal} {Phys. Rev. A}\ }\textbf {\bibinfo {volume} {89}},\ \bibinfo
  {pages} {042120} (\bibinfo {year} {2014})}\BibitemShut {NoStop}%
\bibitem [{\citenamefont {de~Vega}\ and\ \citenamefont
  {Alonso}(2017)}]{deVega2017}%
  \BibitemOpen
  \bibfield  {author} {\bibinfo {author} {\bibfnamefont {I.}~\bibnamefont
  {de~Vega}}\ and\ \bibinfo {author} {\bibfnamefont {D.}~\bibnamefont
  {Alonso}},\ }\href {https://doi.org/10.1103/RevModPhys.89.015001} {\bibfield
  {journal} {\bibinfo  {journal} {Rev. Mod. Phys.}\ }\textbf {\bibinfo {volume}
  {89}},\ \bibinfo {pages} {015001} (\bibinfo {year} {2017})}\BibitemShut
  {NoStop}%
\bibitem [{\citenamefont {Chruściński}\ and\ \citenamefont
  {Kossakowski}(2012)}]{Chruscinski2012}%
  \BibitemOpen
  \bibfield  {author} {\bibinfo {author} {\bibfnamefont {D.}~\bibnamefont
  {Chruściński}}\ and\ \bibinfo {author} {\bibfnamefont {A.}~\bibnamefont
  {Kossakowski}},\ }\href {https://doi.org/10.1088/0953-4075/45/15/154002}
  {\bibfield  {journal} {\bibinfo  {journal} {Journal of Physics B: Atomic,
  Molecular and Optical Physics}\ }\textbf {\bibinfo {volume} {45}},\ \bibinfo
  {pages} {154002} (\bibinfo {year} {2012})}\BibitemShut {NoStop}%
\bibitem [{\citenamefont {Lindblad}(1976)}]{Lindblad1976}%
  \BibitemOpen
  \bibfield  {author} {\bibinfo {author} {\bibfnamefont {G.}~\bibnamefont
  {Lindblad}},\ }\href {https://doi.org/10.1007/bf01608499} {\bibfield
  {journal} {\bibinfo  {journal} {Communications in Mathematical Physics}\
  }\textbf {\bibinfo {volume} {48}},\ \bibinfo {pages} {119} (\bibinfo {year}
  {1976})}\BibitemShut {NoStop}%
\bibitem [{\citenamefont {Gorini}\ \emph {et~al.}(1976)\citenamefont {Gorini},
  \citenamefont {Kossakowski},\ and\ \citenamefont {Sudarshan}}]{GKS}%
  \BibitemOpen
  \bibfield  {author} {\bibinfo {author} {\bibfnamefont {V.}~\bibnamefont
  {Gorini}}, \bibinfo {author} {\bibfnamefont {A.}~\bibnamefont
  {Kossakowski}},\ and\ \bibinfo {author} {\bibfnamefont {E.~C.~G.}\
  \bibnamefont {Sudarshan}},\ }\href {https://doi.org/10.1063/1.522979}
  {\bibfield  {journal} {\bibinfo  {journal} {Journal of Mathematical Physics}\
  }\textbf {\bibinfo {volume} {17}},\ \bibinfo {pages} {821} (\bibinfo {year}
  {1976})}\BibitemShut {NoStop}%
\bibitem [{\citenamefont {Ángel Rivas}\ \emph {et~al.}(2014)\citenamefont
  {Ángel Rivas}, \citenamefont {Huelga},\ and\ \citenamefont
  {Plenio}}]{Rivas2014}%
  \BibitemOpen
  \bibfield  {author} {\bibinfo {author} {\bibnamefont {Ángel Rivas}},
  \bibinfo {author} {\bibfnamefont {S.~F.}\ \bibnamefont {Huelga}},\ and\
  \bibinfo {author} {\bibfnamefont {M.~B.}\ \bibnamefont {Plenio}},\ }\href
  {https://doi.org/10.1088/0034-4885/77/9/094001} {\bibfield  {journal}
  {\bibinfo  {journal} {Reports on Progress in Physics}\ }\textbf {\bibinfo
  {volume} {77}},\ \bibinfo {pages} {094001} (\bibinfo {year}
  {2014})}\BibitemShut {NoStop}%
\bibitem [{\citenamefont {Burkard}\ \emph {et~al.}(2004)\citenamefont
  {Burkard}, \citenamefont {Koch},\ and\ \citenamefont
  {DiVincenzo}}]{Burkard2004}%
  \BibitemOpen
  \bibfield  {author} {\bibinfo {author} {\bibfnamefont {G.}~\bibnamefont
  {Burkard}}, \bibinfo {author} {\bibfnamefont {R.~H.}\ \bibnamefont {Koch}},\
  and\ \bibinfo {author} {\bibfnamefont {D.~P.}\ \bibnamefont {DiVincenzo}},\
  }\href {https://doi.org/10.1103/PhysRevB.69.064503} {\bibfield  {journal}
  {\bibinfo  {journal} {Phys. Rev. B}\ }\textbf {\bibinfo {volume} {69}},\
  \bibinfo {pages} {064503} (\bibinfo {year} {2004})}\BibitemShut {NoStop}%
\bibitem [{\citenamefont {Burkard}(2005)}]{Burkard2005}%
  \BibitemOpen
  \bibfield  {author} {\bibinfo {author} {\bibfnamefont {G.}~\bibnamefont
  {Burkard}},\ }\href {https://doi.org/10.1103/PhysRevB.71.144511} {\bibfield
  {journal} {\bibinfo  {journal} {Phys. Rev. B}\ }\textbf {\bibinfo {volume}
  {71}},\ \bibinfo {pages} {144511} (\bibinfo {year} {2005})}\BibitemShut
  {NoStop}%
\bibitem [{\citenamefont {Solgun}\ \emph {et~al.}(2014)\citenamefont {Solgun},
  \citenamefont {Abraham},\ and\ \citenamefont {DiVincenzo}}]{Solgun2014}%
  \BibitemOpen
  \bibfield  {author} {\bibinfo {author} {\bibfnamefont {F.}~\bibnamefont
  {Solgun}}, \bibinfo {author} {\bibfnamefont {D.~W.}\ \bibnamefont
  {Abraham}},\ and\ \bibinfo {author} {\bibfnamefont {D.~P.}\ \bibnamefont
  {DiVincenzo}},\ }\href {https://doi.org/10.1103/PhysRevB.90.134504}
  {\bibfield  {journal} {\bibinfo  {journal} {Phys. Rev. B}\ }\textbf {\bibinfo
  {volume} {90}},\ \bibinfo {pages} {134504} (\bibinfo {year}
  {2014})}\BibitemShut {NoStop}%
\bibitem [{\citenamefont {Solgun}\ and\ \citenamefont
  {DiVincenzo}(2015)}]{Solgun2015}%
  \BibitemOpen
  \bibfield  {author} {\bibinfo {author} {\bibfnamefont {F.}~\bibnamefont
  {Solgun}}\ and\ \bibinfo {author} {\bibfnamefont {D.~P.}\ \bibnamefont
  {DiVincenzo}},\ }\href
  {https://doi.org/https://doi.org/10.1016/j.aop.2015.07.005} {\bibfield
  {journal} {\bibinfo  {journal} {Annals of Physics}\ }\textbf {\bibinfo
  {volume} {361}},\ \bibinfo {pages} {605} (\bibinfo {year}
  {2015})}\BibitemShut {NoStop}%
\bibitem [{\citenamefont {Parra-Rodriguez}\ \emph {et~al.}(2019)\citenamefont
  {Parra-Rodriguez}, \citenamefont {Egusquiza}, \citenamefont {DiVincenzo},\
  and\ \citenamefont {Solano}}]{Parra2019}%
  \BibitemOpen
  \bibfield  {author} {\bibinfo {author} {\bibfnamefont {A.}~\bibnamefont
  {Parra-Rodriguez}}, \bibinfo {author} {\bibfnamefont {I.~L.}\ \bibnamefont
  {Egusquiza}}, \bibinfo {author} {\bibfnamefont {D.~P.}\ \bibnamefont
  {DiVincenzo}},\ and\ \bibinfo {author} {\bibfnamefont {E.}~\bibnamefont
  {Solano}},\ }\href {https://doi.org/10.1103/PhysRevB.99.014514} {\bibfield
  {journal} {\bibinfo  {journal} {Phys. Rev. B}\ }\textbf {\bibinfo {volume}
  {99}},\ \bibinfo {pages} {014514} (\bibinfo {year} {2019})}\BibitemShut
  {NoStop}%
\bibitem [{\citenamefont {Riwar}\ and\ \citenamefont
  {DiVincenzo}(2022)}]{Riwar2022}%
  \BibitemOpen
  \bibfield  {author} {\bibinfo {author} {\bibfnamefont {R.-P.}\ \bibnamefont
  {Riwar}}\ and\ \bibinfo {author} {\bibfnamefont {D.~P.}\ \bibnamefont
  {DiVincenzo}},\ }\href {https://doi.org/10.1038/s41534-022-00539-x}
  {\bibfield  {journal} {\bibinfo  {journal} {npj Quantum Information}\
  }\textbf {\bibinfo {volume} {8}},\ \bibinfo {pages} {36} (\bibinfo {year}
  {2022})}\BibitemShut {NoStop}%
\bibitem [{\citenamefont {Caldeira}\ and\ \citenamefont {Leggett}(1983)}]{CL}%
  \BibitemOpen
  \bibfield  {author} {\bibinfo {author} {\bibfnamefont {A.}~\bibnamefont
  {Caldeira}}\ and\ \bibinfo {author} {\bibfnamefont {A.}~\bibnamefont
  {Leggett}},\ }\href {https://doi.org/10.1016/0003-4916(83)90202-6} {\bibfield
   {journal} {\bibinfo  {journal} {Annals of Physics}\ }\textbf {\bibinfo
  {volume} {149}},\ \bibinfo {pages} {374} (\bibinfo {year}
  {1983})}\BibitemShut {NoStop}%
\bibitem [{\citenamefont {Clerk}\ \emph {et~al.}(2010)\citenamefont {Clerk},
  \citenamefont {Devoret}, \citenamefont {Girvin}, \citenamefont {Marquardt},\
  and\ \citenamefont {Schoelkopf}}]{Noise2010}%
  \BibitemOpen
  \bibfield  {author} {\bibinfo {author} {\bibfnamefont {A.~A.}\ \bibnamefont
  {Clerk}}, \bibinfo {author} {\bibfnamefont {M.~H.}\ \bibnamefont {Devoret}},
  \bibinfo {author} {\bibfnamefont {S.~M.}\ \bibnamefont {Girvin}}, \bibinfo
  {author} {\bibfnamefont {F.}~\bibnamefont {Marquardt}},\ and\ \bibinfo
  {author} {\bibfnamefont {R.~J.}\ \bibnamefont {Schoelkopf}},\ }\href
  {https://doi.org/10.1103/RevModPhys.82.1155} {\bibfield  {journal} {\bibinfo
  {journal} {Rev. Mod. Phys.}\ }\textbf {\bibinfo {volume} {82}},\ \bibinfo
  {pages} {1155} (\bibinfo {year} {2010})}\BibitemShut {NoStop}%
\bibitem [{\citenamefont {Breuer}\ \emph {et~al.}(2004)\citenamefont {Breuer},
  \citenamefont {Burgarth},\ and\ \citenamefont {Petruccione}}]{Breuer2004}%
  \BibitemOpen
  \bibfield  {author} {\bibinfo {author} {\bibfnamefont {H.-P.}\ \bibnamefont
  {Breuer}}, \bibinfo {author} {\bibfnamefont {D.}~\bibnamefont {Burgarth}},\
  and\ \bibinfo {author} {\bibfnamefont {F.}~\bibnamefont {Petruccione}},\
  }\href {https://doi.org/10.1103/PhysRevB.70.045323} {\bibfield  {journal}
  {\bibinfo  {journal} {Phys. Rev. B}\ }\textbf {\bibinfo {volume} {70}},\
  \bibinfo {pages} {045323} (\bibinfo {year} {2004})}\BibitemShut {NoStop}%
\bibitem [{\citenamefont {Nakajima}(1958)}]{Nakajima1958}%
  \BibitemOpen
  \bibfield  {author} {\bibinfo {author} {\bibfnamefont {S.}~\bibnamefont
  {Nakajima}},\ }\href {https://doi.org/10.1143/ptp.20.948} {\bibfield
  {journal} {\bibinfo  {journal} {Progress of Theoretical Physics}\ }\textbf
  {\bibinfo {volume} {20}},\ \bibinfo {pages} {948} (\bibinfo {year}
  {1958})}\BibitemShut {NoStop}%
\bibitem [{\citenamefont {Zwanzig}(1960)}]{Zwanzig1960}%
  \BibitemOpen
  \bibfield  {author} {\bibinfo {author} {\bibfnamefont {R.}~\bibnamefont
  {Zwanzig}},\ }\href {https://doi.org/10.1063/1.1731409} {\bibfield  {journal}
  {\bibinfo  {journal} {The Journal of Chemical Physics}\ }\textbf {\bibinfo
  {volume} {33}},\ \bibinfo {pages} {1338} (\bibinfo {year}
  {1960})}\BibitemShut {NoStop}%
\bibitem [{\citenamefont {Chaturvedi}\ and\ \citenamefont
  {Shibata}(1979)}]{Chaturvedi1979}%
  \BibitemOpen
  \bibfield  {author} {\bibinfo {author} {\bibfnamefont {S.}~\bibnamefont
  {Chaturvedi}}\ and\ \bibinfo {author} {\bibfnamefont {F.}~\bibnamefont
  {Shibata}},\ }\href {https://doi.org/10.1007/bf01319852} {\bibfield
  {journal} {\bibinfo  {journal} {Zeitschrift für Physik B Condensed Matter
  and Quanta}\ }\textbf {\bibinfo {volume} {35}},\ \bibinfo {pages} {297}
  (\bibinfo {year} {1979})}\BibitemShut {NoStop}%
\bibitem [{\citenamefont {Breuer}\ \emph {et~al.}(2001)\citenamefont {Breuer},
  \citenamefont {Kappler},\ and\ \citenamefont {Petruccione}}]{TCL1}%
  \BibitemOpen
  \bibfield  {author} {\bibinfo {author} {\bibfnamefont {H.-P.}\ \bibnamefont
  {Breuer}}, \bibinfo {author} {\bibfnamefont {B.}~\bibnamefont {Kappler}},\
  and\ \bibinfo {author} {\bibfnamefont {F.}~\bibnamefont {Petruccione}},\
  }\href {https://doi.org/https://doi.org/10.1006/aphy.2001.6152} {\bibfield
  {journal} {\bibinfo  {journal} {Annals of Physics}\ }\textbf {\bibinfo
  {volume} {291}},\ \bibinfo {pages} {36} (\bibinfo {year} {2001})}\BibitemShut
  {NoStop}%
\bibitem [{\citenamefont {Kubo}(1963)}]{Kubo1963}%
  \BibitemOpen
  \bibfield  {author} {\bibinfo {author} {\bibfnamefont {R.}~\bibnamefont
  {Kubo}},\ }\href {https://doi.org/10.1063/1.1703941} {\bibfield  {journal}
  {\bibinfo  {journal} {Journal of Mathematical Physics}\ }\textbf {\bibinfo
  {volume} {4}},\ \bibinfo {pages} {174} (\bibinfo {year} {1963})}\BibitemShut
  {NoStop}%
\bibitem [{\citenamefont {Megier}\ \emph {et~al.}(2017)\citenamefont {Megier},
  \citenamefont {Chru{\'{s}}ci{\'{n}}ski}, \citenamefont {Piilo},\ and\
  \citenamefont {Strunz}}]{Megier2017}%
  \BibitemOpen
  \bibfield  {author} {\bibinfo {author} {\bibfnamefont {N.}~\bibnamefont
  {Megier}}, \bibinfo {author} {\bibfnamefont {D.}~\bibnamefont
  {Chru{\'{s}}ci{\'{n}}ski}}, \bibinfo {author} {\bibfnamefont
  {J.}~\bibnamefont {Piilo}},\ and\ \bibinfo {author} {\bibfnamefont {W.~T.}\
  \bibnamefont {Strunz}},\ }\href {https://doi.org/10.1038/s41598-017-06059-5}
  {\bibfield  {journal} {\bibinfo  {journal} {Scientific Reports}\ }\textbf
  {\bibinfo {volume} {7}},\ \bibinfo {pages} {6379} (\bibinfo {year}
  {2017})}\BibitemShut {NoStop}%
\bibitem [{\citenamefont {Andersson}\ \emph {et~al.}(2007)\citenamefont
  {Andersson}, \citenamefont {Cresser},\ and\ \citenamefont
  {Hall}}]{Andersson2007}%
  \BibitemOpen
  \bibfield  {author} {\bibinfo {author} {\bibfnamefont {E.}~\bibnamefont
  {Andersson}}, \bibinfo {author} {\bibfnamefont {J.~D.}\ \bibnamefont
  {Cresser}},\ and\ \bibinfo {author} {\bibfnamefont {M.~J.~W.}\ \bibnamefont
  {Hall}},\ }\href {https://doi.org/10.1080/09500340701352581} {\bibfield
  {journal} {\bibinfo  {journal} {Journal of Modern Optics}\ }\textbf {\bibinfo
  {volume} {54}},\ \bibinfo {pages} {1695} (\bibinfo {year}
  {2007})}\BibitemShut {NoStop}%
\bibitem [{\citenamefont {Choi}(1975)}]{Choi1975}%
  \BibitemOpen
  \bibfield  {author} {\bibinfo {author} {\bibfnamefont {M.-D.}\ \bibnamefont
  {Choi}},\ }\href {https://doi.org/10.1016/0024-3795(75)90075-0} {\bibfield
  {journal} {\bibinfo  {journal} {Linear Algebra and its Applications}\
  }\textbf {\bibinfo {volume} {10}},\ \bibinfo {pages} {285} (\bibinfo {year}
  {1975})}\BibitemShut {NoStop}%
\bibitem [{\citenamefont {M\"akel\"a}\ and\ \citenamefont
  {M\"ott\"onen}(2013)}]{RWAnm}%
  \BibitemOpen
  \bibfield  {author} {\bibinfo {author} {\bibfnamefont {H.}~\bibnamefont
  {M\"akel\"a}}\ and\ \bibinfo {author} {\bibfnamefont {M.}~\bibnamefont
  {M\"ott\"onen}},\ }\href {https://doi.org/10.1103/PhysRevA.88.052111}
  {\bibfield  {journal} {\bibinfo  {journal} {Phys. Rev. A}\ }\textbf {\bibinfo
  {volume} {88}},\ \bibinfo {pages} {052111} (\bibinfo {year}
  {2013})}\BibitemShut {NoStop}%
\bibitem [{\citenamefont {Kraus}\ \emph {et~al.}(1983)\citenamefont {Kraus},
  \citenamefont {B\"{o}hm}, \citenamefont {Dollard},\ and\ \citenamefont
  {Wootters}}]{Kraus1983}%
  \BibitemOpen
  \bibinfo {editor} {\bibfnamefont {K.}~\bibnamefont {Kraus}}, \bibinfo
  {editor} {\bibfnamefont {A.}~\bibnamefont {B\"{o}hm}}, \bibinfo {editor}
  {\bibfnamefont {J.~D.}\ \bibnamefont {Dollard}},\ and\ \bibinfo {editor}
  {\bibfnamefont {W.~H.}\ \bibnamefont {Wootters}},\ eds.,\ \href
  {https://doi.org/10.1007/3-540-12732-1} {\emph {\bibinfo {title} {States,
  Effects, and Operations Fundamental Notions of Quantum Theory}}}\ (\bibinfo
  {publisher} {Springer Berlin Heidelberg},\ \bibinfo {year}
  {1983})\BibitemShut {NoStop}%
\bibitem [{\citenamefont {Krantz}\ \emph {et~al.}(2019)\citenamefont {Krantz},
  \citenamefont {Kjaergaard}, \citenamefont {Yan}, \citenamefont {Orlando},
  \citenamefont {Gustavsson},\ and\ \citenamefont {Oliver}}]{Krantz2019}%
  \BibitemOpen
  \bibfield  {author} {\bibinfo {author} {\bibfnamefont {P.}~\bibnamefont
  {Krantz}}, \bibinfo {author} {\bibfnamefont {M.}~\bibnamefont {Kjaergaard}},
  \bibinfo {author} {\bibfnamefont {F.}~\bibnamefont {Yan}}, \bibinfo {author}
  {\bibfnamefont {T.~P.}\ \bibnamefont {Orlando}}, \bibinfo {author}
  {\bibfnamefont {S.}~\bibnamefont {Gustavsson}},\ and\ \bibinfo {author}
  {\bibfnamefont {W.~D.}\ \bibnamefont {Oliver}},\ }\href
  {https://doi.org/10.1063/1.5089550} {\bibfield  {journal} {\bibinfo
  {journal} {Applied Physics Reviews}\ }\textbf {\bibinfo {volume} {6}},\
  \bibinfo {pages} {021318} (\bibinfo {year} {2019})}\BibitemShut {NoStop}%
\end{thebibliography}%

\end{document}